\documentclass[hyper,aps,eqsecnum,superscriptaddress,12pt]{revtex4}
\usepackage{bbm}
\usepackage{mathrsfs}
\usepackage{amsmath,amssymb,bm,comment}
\usepackage{hyperref}
\usepackage{graphicx}
\usepackage{dcolumn}
\usepackage{bm}
\usepackage{stmaryrd}
\usepackage{color}
\usepackage{fancyhdr}
\begin{document}

\title{ Second-order Charge Currents and Stress Tensor in Chiral System }

\author{Shi-Zheng Yang}
\email{yangshizheng@mail.sdu.edu.cn}
\affiliation{Key Laboratory of Particle Physics and Particle Irradiation (MOE),
Institute of Frontier and Interdisciplinary Science,
Shandong University, Qingdao, Shandong 266237, China}

\author{Jian-Hua Gao}
\email{gaojh@sdu.edu.cn}
\affiliation{Shandong Provincial Key Laboratory of Optical Astronomy and Solar-Terrestrial Environment,
Institute of Space Sciences, Shandong University, Weihai, Shandong 264209, China}

\author{Zuo-Tang Liang}
\email{liang@sdu.edu.cn}
\affiliation{Key Laboratory of Particle Physics and Particle Irradiation (MOE),
Institute of Frontier and Interdisciplinary Science,
Shandong University, Qingdao, Shandong 266237, China}

\author{Qun Wang}
\email{qunwang@ustc.edu.cn}
\affiliation{Interdisciplinary Center for Theoretical Study and Department of Modern Physics, University of Science and Technology
of China, Hefei, Anhui 230026, China}
\affiliation{Peng Huanwu Center for Fundamental Theory, Hefei, Anhui 230026, China}

\begin{abstract}
We study Wigner equations for massless spin-1/2 charged fermions at global equilibrium in static and uniform vorticity and electromagnetic fields.
The Wigner functions can be solved order by order from Wigner equations through  the power expansion of the vorticity and electromagnetic fields.
The non-dissipative charge currents and the stress
tensor up to  the second order can be obtained from Wigner functions. The charge and energy densities and the pressure have contributions from  vorticity
and electromagnetic  fields at the second order. The vector and axial Hall currents can be induced along the direction
orthogonal to vorticity and electromagnetic fields at the second order. We also find that the trace anomaly emerges
 naturally in renormalization of the stress tensor by including  quantum corrections from  electromagnetic fields.
\end{abstract}

\maketitle
\thispagestyle{fancy}
\renewcommand{\headrulewidth}{0pt}
\section{Introduction}
\label{sec:intro}

It is well-known in classical electrodynamics that the electromagnetic
field can generate electric currents, such as Olm's current from electric fields or Hall's current
from magnetic fields. There are also currents from quantum effects which attract broad interest
in  high energy nuclear physics and condensed matter physics.
One example is  chiral anomaly, a pure quantum effect, in which currents
along the external magnetic field can be induced, it is called the chiral magnetic effect (CME)
\cite{Vilenkin:1980fu,Kharzeev:2007jp,Fukushima:2008xe} . The vorticity  in an ideal fluid behaves like a magnetic field.
Similar to CME, the vorticity  can induce the electric current
in a charged fluid of massless fermions, which is called the chiral vortical effect (CVE) \cite{Vilenkin:1978hb,Kharzeev:2007tn,Erdmenger:2008rm,Banerjee:2008th}.
In addition to CME and CVE, chiral currents can also be generated by vorticity and magnetic fields, these are called the chiral separate effect (CSE) \cite{Son:2004tq,Metlitski:2005pr}
or the local polarization effect (LPE) \cite{Gao:2012ix}.
Theoretical studies of these effects have been carried out within a variety
of approaches, such as AdS/CFT duality
\cite{Newman:2005hd,Yee:2009vw,Rebhan:2009vc,Gorsky:2010xu,Gynther:2010ed,Hoyos:2011us,Amado:2011zx,
Nair:2011mk,Kalaydzhyan:2011vx,Lin:2013sga}, relativistic hydrodynamics
\cite{Son:2009tf,Sadofyev:2010pr,Pu:2010as,Kharzeev:2011ds}, quantum field theory
\cite{Kharzeev:2007jp,Fukushima:2008xe,Kharzeev:2009pj,Fukushima:2009ft,Asakawa:2010bu,Fukushima:2010vw,
Fukushima:2010zza,Landsteiner:2011cp,Hou:2011ze,Hou:2012xg,Lin:2018aon,Feng:2018tpb,Dong:2020zci} and chiral kinetic theories
\cite{Gao:2012ix,Stephanov:2012ki,Son:2012zy,Chen:2012ca,Manuel:2013zaa,Chen:2014cla,Chen:2015gta,Hidaka:2016yjf,Mueller:2017lzw,Huang:2017tsq,
Huang:2018wdl,Hidaka:2018ekt,Gao:2018wmr,Gao:2018jsi,Liu:2018xip}.

From the point of view of hydrodynamics, these anomalous currents are non-dissipative
without entropy production and they all appear at the first order in space-time derivatives.
It has been shown \cite{Muller:1967zza,Israel:1976tn,Israel:1979wp,Hiscock:1983zz,Hiscock:1985zz,Hiscock:1987zz}
that the relativistic hydrodynamical equations with only first order term are acausal and unstable.
This issue can be repaired by including  second-order terms.  We also need to include  higher order contributions
when  vorticity or  electromagnetic fields are strong enough.
This is the case in high energy heavy ion collisions, in which both strong magnetic fields
\cite{Bzdak:2011yy,Deng:2012pc,Bloczynski:2012en} and vorticity fields \cite{Liang:2004ph,Gao:2007bc,Becattini:2007sr,
Csernai:2013bqa,Jiang:2016woz,Deng:2016gyh,Pang:2016igs} are generated in non-central collisions.
There have been already some earlier attempts to study transport phenomena at the second order
 in chiral systems including second order hydrodynamics with reversal invariance \cite{Kharzeev:2011ds}, Kubo formula or diagrammatic methods from the quantum field theory \cite{Jimenez-Alba:2015bia,Hattori:2016njk,Buzzegoli:2017cqy,Buzzegoli:2018wpy},  chiral kinetic theories \cite{Satow:2014lia,Gorbar:2017cwv,Gorbar:2017toh,Abbasi:2018zoc},
and equilibrium partition functions or AdS/CFT dualities \cite{Banerjee:2012iz,Bhattacharyya:2013ida,Megias:2014mba,Bu:2019qmd}.

The Lorentz covariant and gauge invariant quantum transport theories  ~\cite{Heinz:1983nx,Elze:1986qd,Vasak:1987um,Zhuang:1995pd} based
on Wigner functions can be derived from quantum field theory and are
expected include all quantum corrections.
In  previous works \cite{Gao:2012ix,Gao:2015zka} by some of us,
a power expansion  in space-time derivatives and weak fields for
Wigner functions of chiral fermions was proposed  near equilibrium.
It turns out that the Wigner function formalism is successful to
reproduce first order currents in CME, CVE, CSE and LPE.
In this paper, we will use the power expansion method to
derive  second order non-dissipative charge currents
and energy-momentum tensor in a non-interacting fluid.
The novelty of non-dissipative currents   is that they are present  in global equilibrium.
This provides a unique shortcut to investigate these effects because the calculation will be greatly simplified in global equilibrium. However in order to arrive at non-trivial global equilibrium, we must choose static and uniform vorticity and electromagnetic fields. Besides, we will neglect the fermion mass (chiral limit) and make
chiral limit,  which is valid when the temperature is much greater than the particle's static mass.

In Sec.~\ref{sec:Wigner}, we give a brief overview of the Wigner function formalism
for a chiral fermion system. In Sec.~\ref{sec:solve}, we solve
the equations for the covariant Wigner function {  at global equilibrium  with static and uniform  vorticity and electromagnetic fields}
 by using the method of
Refs.\cite{Gao:2012ix,Gao:2015zka,Gao:2017gfq,Gao:2018jsi}.
We give the { simplest}  solution to the Wigner function up to the second order of the vorticity and
electromagnetic field. In Sec.~\ref{sec:current} and Sec.~\ref{sec:tensor},
we present the induced vector and axial currents
and energy-momentum tensor up to the second order.
It can be verified that the charge (vector current) conservation  and the anomalous conservation of axial current hold automatically.
There is no infrared and ultraviolet  divergence for the vector and axial charges. For the energy-momentum tensor at the second order,
the contribution  from the  vorticity only and that from  the vorticity and electromagnetic field are both finite, while
the contribution from the electromagnetic field has logarithmic ultraviolet divergence when the Dirac sea or vacuum contribution is included.
With a proper dimension regularization, we  obtain the  results that satisfy the energy-momentum conservation.
Especially, after we renormalize the stress tensor by including the quantum correction from the electromagnetic field,
the trace anomaly emerges naturally. In  Sec.~\ref{sec:cons}, we verify  the conservation of the electric charge and the energy momentum as well as
the anomalous conservation of the axial charge. In Sec.~\ref{sec:cons}, we extend the special solution to general ones and
compare  with previous results obtained in Ref \cite{Buzzegoli:2017cqy,Buzzegoli:2018wpy} .
A summary of our results is made in Sec.~\ref{sec:summary}.

 We use the convention for the metric tensor $g_{\mu\nu}=\mathrm{diag}(1,-1,-1,-1)$
and the Levi-Civita tensor $\epsilon^{0123}=1$. For notational simplicity, the electric charge of the fermion is absorbed into
the vector potential $A^\mu$.


\section{Wigner function formalism}
\label{sec:Wigner}

The Wigner function $W(x,p)$ for Dirac fermions is a $4\times 4$ matrix and is defined as
the ensemble average of the Wigner operator \cite{Heinz:1983nx,Elze:1986qd,Vasak:1987um},
\begin{equation}
\label{wigner}
W_{\alpha\beta}(x,p) = \int\frac{d^4 y}{(2\pi)^4}
e^{-ip\cdot y}\left\langle \bar\psi_\beta\left(x+\frac{y}{2}\right)
U\left(x+\frac{y}{2},x-\frac{y}{2}\right)
\psi_{\alpha}\left(x-\frac{y}{2}\right)\right\rangle,
\end{equation}
where $U$ denotes the gauge link along the straight line between $x-y/2$ and $x+y/2$,
\begin{equation}
\label{link}
U\left(x+\frac{y}{2},x-\frac{y}{2}\right)  \equiv \textrm{Exp} \left(-i \int_{x-y/2}^{x+y/2} dz^\mu A_\mu (z)\right).
\end{equation}
We will  restrict ourselves to a system of chiral fermions without collisions in a constant
external electromagnetic field $F^{\mu\nu}$ in space and time, i.e. $\partial^\lambda F^{\mu\nu}=0$, hence we have
removed the path ordering of the gauge link.
The Wigner equation for chiral fermions in a constant electromagnetic field  is given by \cite{Vasak:1987um},
\begin{equation}
\label{eq-c}
\gamma_\mu \left( p^\mu +\frac{i}{2}\nabla^\mu\right)  W(x,p)=0 ,
\end{equation}
where $\gamma^{\mu}$ are Dirac matrices and
$\nabla^\mu \equiv \partial^\mu_x - F^{\mu\nu}\partial_\nu^p$ with $\partial_x\;(\partial^p)$
being the derivative with respect to $x\;(p)$.
Since the Wigner equation is derived from the Dirac equation, the bilinear operator in the  Wigner function should not be  normal ordered.
It has been demonstrated in Ref. \cite{Gao:2019zhk} that this feature plays a central role to give rise to the chiral anomaly in quantum kinetic theory.
We can decompose the Wigner function in terms of 16 independent generators of the Clifford algebra,
\begin{eqnarray}
\label{decomposition}
W&=& \frac{1}{4}\left[\mathscr{F}+i\gamma^5 \mathscr{P}
+\gamma^\mu \mathscr{V}_\mu   +\gamma^5 \gamma^\mu \mathscr{A}_\mu
+\frac{1}{2}\sigma^{\mu\nu} \mathscr{S}_{\mu\nu}\right]\;,
\end{eqnarray}
where we have suppressed arguments of the Wigner function for notational simplicity.

For chiral fermions, it is more convenient to define the chiral component
\begin{eqnarray}
\label{chibasis}
\mathscr{J}_s^\mu\equiv\frac{1}{2}\left(\mathscr{V}^\mu +s \mathscr{A}^\mu\right)\;,
\end{eqnarray}
with $s=+ 1$ and $-1$ corresponding to the right-hand and left-hand component respectively.
Substituting  Eq. (\ref{decomposition}) and Eq. (\ref{chibasis}) into Eq.(\ref{eq-c}),
we find that the right-hand  or left-hand component are decoupled from other components and satisfy
\begin{eqnarray}
\label{Js-eq}
\nabla_\mu\mathscr{J}_s^\mu &=& 0,\\
\label{Js-c1}
p_\mu\mathscr{J}_s^\mu &=&0,\\
\label{Js-c2}
p_\mu \mathscr{J}_{s\nu}-p_\nu \mathscr{J}_{s\mu} &=&-\frac{s}{2} \epsilon_{\mu\nu\rho\sigma}\nabla^\rho \mathscr{J}_s^\sigma.
\end{eqnarray}
We will suppress the subscript $s$ in Sec.~\ref{sec:solve} for notational simplicity and recover it in Sec.~\ref{sec:current}.

\section{Wigner function near equilibrium}
\label{sec:solve}

We assume that both the space-time derivative $\partial_x$ and the field strength
$F^{\mu\nu}$ in the operator $\nabla_\mu$ are small variables of the same order and
 play the role of expansion parameters.
We solve the Wigner equation by the covariant perturbation method developed
in Refs. \cite{Gao:2012ix,Gao:2018jsi,Gao:2017gfq} and present the solution
near equilibrium up to the second order in $\partial_x$ and $F^{\mu\nu}$.
In fact, this expansion is equivalent to an expansion in the Planck constant $\hbar$
(or the semiclassical expansion) because $\hbar$ always comes with $\nabla_\mu$.
According to the perturbation method, the Wigner function can be obtained order by order,
\begin{equation}
\mathscr{J}_\mu = \mathscr{J}_\mu^{(0)} + \mathscr{J}_\mu^{(1)} +  \mathscr{J}_\mu^{(2)} + \cdots \;,
\end{equation}
where the superscripts $(0),(1),...$ denote the orders of the power in the expansion.
Substituting this expansion into Wigner equations from (\ref{Js-eq})
to (\ref{Js-c2}) and requiring that the equations hold order by order.
The equations for $\mathscr{J}_\mu^{(n)}$ with $n > 0$ read
\begin{eqnarray}
\label{Js-eq-n}
\nabla_\mu\mathscr{J}^{(n)\mu} &=& 0,\\
\label{Js-c1-n}
p_\mu\mathscr{J}^{(n)\mu} &=&0,\\
\label{Js-c2-n}
p_\mu \mathscr{J}^{(n)}_{\nu}-p_\nu \mathscr{J}^{(n)}_{\mu}&=&-\frac{s}{2} \epsilon_{\mu\nu\rho\sigma}\nabla^\rho \mathscr{J}^{(n-1)\sigma}.
\end{eqnarray}
If we define $\mathscr{J}^{(-1)\sigma}=0$, Eq. (\ref{Js-c2-n}) also works for $n=0$.
When we contract both sides of Eq. (\ref{Js-c2-n}) with $p^\nu$, we have
\begin{eqnarray}
\label{J0-n}
p^2 \mathscr{J}_\mu^{(n)} &=&\frac{s}{2}\epsilon_{\mu\nu\rho\sigma}p^\nu \nabla^\rho \mathscr{J}^{(n-1)\sigma}\;,
\end{eqnarray}
where we have used Eq. (\ref{Js-c1-n}). Hence the general form of $\mathscr{J}_\mu^{(n)}$ is
\begin{eqnarray}
\label{Jmu-n}
\mathscr{J}_\mu^{(n)} &=& {J}_\mu^{(n)} \delta(p^2) + \frac{s}{2p^2}\epsilon_{\mu\nu\rho\sigma}p^\nu
\nabla^\rho \mathscr{J}^{(n-1)\sigma} \;,
\end{eqnarray}
where ${J}_\mu^{(n)}$ is nonsingular at $p^2=0$.
This expression is an iterative equation connecting the $n$-th order solution
with the $(n-1)$-th order one. The constraint condition (\ref{Js-c1-n}) gives
\begin{eqnarray}
\label{J-c1-onshell}
p^\mu {J}_\mu^{(n)} \delta(p^2) =0.
\end{eqnarray}
In general, we can decompose ${J}_\mu^{(n)}$ into two parts
\begin{eqnarray}
\label{calJ}
{J}_\mu^{(n)}(x,p)=p_\mu f^{(n)}(x,p) +{X}_\mu^{(n)}(x,p)\;,
\end{eqnarray}
where the first term  satisfies Eq.~(\ref{J-c1-onshell}) automatically due to $p^2\delta(p^2)=0$
and the second term is assumed to satisfy $p^\mu{X}_\mu^{(n)}=0$ when there is no mass-shell constraint.


It is straightforward to write down the zeroth order solution,
\begin{eqnarray}
\label{Jmu-0}
\mathscr{J}_\mu^{(0)}(x,p) &=&p_\mu f(x,p) \delta(p^2)\; ,
\end{eqnarray}
without ${X}_\mu^{(0)}$ component. We note that in the above expression we have suppressed
the superscript ${(0)}$ in $f$ because we will set all higher order contributions $f^{(n)}$
for $n\ge 1$ vanish {before  Sec.~\ref{sec:Extend} in which  all possible solutions
for $f^{(1)}$ and $f^{(2)}$ will be discussed. }
Substituting the expression (\ref{Jmu-0}) into  Eq.~(\ref{Js-eq-n}) with $n=0$ gives
the kinetic equation at the zeroth order
\begin{eqnarray}
\label{Js-eq-0}
\delta(p^2)  p^\mu \nabla_\mu  f(x,p)  &=& 0\;.
\end{eqnarray}
Since we try to obtain the solution near equilibrium, at the zeroth order we can choose $f$ as
the Fermi-Dirac distribution function,
\begin{eqnarray}
\label{f-p}
f &=& \frac{1}{4\pi^3}\frac{1}{e^{\beta\cdot p-\bar\mu_s}+1}, \hspace{2cm}   (p_0>0) \\
\label{f-m}
f &=& \frac{1}{4\pi^3} \left(\frac{1}{e^{-\beta\cdot p + \bar \mu_s}+1}-1\right)\;,\;\;\;  (p_0<0)
\end{eqnarray}
where
\begin{eqnarray}
\beta^\mu\equiv \beta u^\mu = \frac{u^\mu}{T},\ \ \bar \mu_s \equiv \frac{\mu_s}{T} = \bar\mu + s \bar \mu_5 ,\ \  \bar\mu=\frac{\mu}{T} ,\ \  \bar\mu_5= \frac{\mu_5}{T} \;,
\end{eqnarray}
with $u$ being the fluid four-velocity, $T$ the temperature,
$\mu_s$ the right-hand/left-hand chemical potential, $\mu$ the vector chemical potential
and $\mu_5$ the axial chemical potential. {We can always introduce
the axial chemical potential in the zeroth-order solution for chiral fermions because the axial current is always
conserved  when  there is no electromagnetic field at the zeroth  order. Actually we can even introduce the axial
chemical potential when the electromagnetic field is present because we can redefine the conserving axial current
by absorbing the Chern-Simons current. The total current is conserved and we can introduce the chemical potential
 corresponding to this conserved charge. }
In the solution given in Eqs.(\ref{f-p}) and (\ref{f-m}),
we see that $\mathscr{J}_\mu^{(0)}(x,p)$ or $f(x,p) $ depends on $x$ only
through $u(x)$, $T(x)$, $\mu(x)$ and $\mu_5(x)$.
The Dirac sea (or vacuum) contribution $-1$ in the anti-particle distribution \cite{Sheng:2017lfu,Sheng:2018jwf}
is indispensable because there is no normal ordering in the definition of the Wigner function (\ref{wigner}).
With the distribution (\ref{f-p}) and (\ref{f-m}), the Wigner function $\mathscr{J}_\mu^{(0)}$
is in the form
\begin{eqnarray}
\mathscr{J}_\mu^{(0)} &=& \frac{p_\mu}{4\pi^3 } \left[
 \frac{1}{e^{\beta\cdot p-\bar\mu_s }+1}\frac{\delta(p_0-|{\bf p}|)}{2 |{\bf p}|}
 +\left( \frac{1}{e^{-\beta\cdot p + \bar\mu_s }+1}-1\right) \frac{\delta(p_0+|{\bf p}|)}{2 |{\bf p}|}\right]\;.
\end{eqnarray}
Inserting Eq.~(\ref{f-p}) or (\ref{f-m}) into the kinetic equation (\ref{Js-eq-0}) we obtain
\begin{eqnarray}
\delta(p^2)  p^\mu  \nabla_{\mu}f  =
 f^{\prime}\left[\frac{1}{2}p^\mu p^{\nu}(\partial_{\mu}\beta_{\nu} + \partial_{\nu}\beta_{\mu})
 -p^\mu \partial_{\mu}\bar{\mu}- p^\mu F_{\mu\nu}\beta^\nu - s p^\mu \partial_{\mu}\bar{\mu}_5 \right]
=0 \;, \label{eq:nabla-f}
\end{eqnarray}
where we have used the shorthand notation $f^{\prime}\equiv {\partial f}/{\partial(\beta\cdot p)}$.
It is obvious that when the constraint conditions
\begin{eqnarray}
\label{pd1}
\partial_\mu \beta_\nu +\partial_\nu\beta_\mu &=& 0,\\
\label{pd3}
\partial_\mu \bar\mu + F_{\mu\nu} \beta^\nu &=& 0 ,\\
\label{pd2}
\partial_\mu\bar\mu_5 &=& 0 \; ,
\end{eqnarray}
are all satisfied, the Wigner function (\ref{Jmu-0}) with (\ref{f-p}) and  (\ref{f-m}) are indeed the solution
to Eq.~(\ref{Js-eq-0}). {These conditions are actually  global equilibrium conditions for the system under static and uniform vorticity and electromagnetic fields.} General solutions to these constraint conditions are
\begin{eqnarray}
\label{beta-mu}
\beta_\mu &=&-\Omega_{\mu\nu}x^\nu \;,\\
\bar\mu &=&-\frac{1}{2}F^{\mu\lambda} x_\lambda \Omega_{\mu \nu} x^\nu + c \;,\\
\bar\mu_5&=& c_5\;,
\end{eqnarray}
together with the integrability condition
\begin{eqnarray}
\label{integrability}
{F_{\lambda}}^\mu \Omega^{\nu\lambda}-{F_{\lambda}}^\nu \Omega^{\mu\lambda}=0\;,
\end{eqnarray}
where $\Omega^{\mu\nu}$ and $c_5/c $ are constant antisymmetric tensor and constants, respectively.
 The integrability condition is obtained by differentiating both sides of Eq.(\ref{pd3}) with $\partial_\nu$  and applying the
commutativity of partial derivatives
\begin{eqnarray}
\label{pd3-a}
\partial_\nu \partial_\mu \bar\mu =\partial_\mu \partial_\nu \bar \mu
\ \ \Longrightarrow\ \   F_{\mu\lambda}\partial_\nu  \beta^\lambda = F_{\nu\lambda}\partial_\mu  \beta^\lambda ,
\end{eqnarray}
which leads to the condition (\ref{integrability}) directly with  Eq.(\ref{beta-mu}).
It should be noted that $\Omega_{\mu\nu}$ is nothing but the
thermal vorticity tensor of the fluid
\begin{eqnarray}
\Omega_{\mu\nu}=\frac{1}{2}\left(\partial_\mu \beta_\nu -\partial_\nu \beta_\mu\right)\;.
\end{eqnarray}


Substituting the zeroth order solution (\ref{Jmu-0}) into Eq.~(\ref{Jmu-n}) with $n=1$
gives rise to the first order solution
\begin{eqnarray}
\label{Jmu-1}
\mathscr{J}_\mu^{(1)}&=& {J}_\mu^{(1)} \delta(p^2) + \frac{s}{2p^2}\epsilon_{\mu\nu\rho\sigma}p^\nu \nabla^\rho \mathscr{J}^{(0)\sigma}\nonumber\\
 &=& {X}_\mu^{(1)} \delta(p^2) + s \tilde F_{\mu\nu}p^\nu  f \delta'(p^2) \;,
\end{eqnarray}
where we have dropped the term  proportional to  $ p_\mu \delta(p^2) $ and used
 $\tilde F^{\mu\nu }=(1/2)\epsilon^{\mu\nu\rho\sigma}F_{\rho\sigma}$ and
$\delta ^\prime (x)= -(1/x)\delta (x)$. The unknown ${X}_\mu^{(1)}$ can be further constrained
by inserting Eq.~(\ref{Jmu-1}) into  Eq.~(\ref{Js-c2-n})
\begin{eqnarray}
\label{X-1}
\left(p_\mu {X}_\nu^{(1)}-p_\nu {X}_\mu^{(1)}\right)\delta(p^2)
&=&\frac{s}{2}\epsilon_{\mu\nu\lambda\rho} p^\lambda \nabla^\rho f\delta(p^2)\nonumber\\
&=&-\frac{s}{2}\left(p_\mu \tilde\Omega_{\nu\lambda}p^\lambda - p_\nu \tilde\Omega_{\mu\lambda}p^\lambda\right)f' \delta(p^2) \;,
\end{eqnarray}
where $\tilde\Omega_{\mu\nu}=(1/2)\epsilon_{\mu\nu\rho\sigma}\Omega^{\rho\sigma}$.
In order to arrive at the last equation, we have
used the specific distribution (\ref{f-p}-\ref{f-m}) and  conditions (\ref{pd1}-\ref{pd2}).
Obviously, from the equation above,  we can set
\begin{eqnarray}
X_\mu^{(1)} =-\frac{s}{2}\tilde\Omega_{\mu\lambda}p^\lambda f' \;,
\end{eqnarray}
which results in
\begin{eqnarray}
\label{Jmu-1-a}
\mathscr{J}^{(1)}_\mu &=&-\frac{s}{2}\tilde\Omega_{\mu\lambda}p^\lambda f' \delta(p^2)
+\frac{s}{2}\epsilon_{\mu\nu\rho\sigma}p^\nu F^{\rho\sigma} f\delta'(p^2) \;.
\end{eqnarray}
Under global equilibrium conditions (\ref{pd1}-\ref{pd2}),
it is straightforward to verify that above $\mathscr{J}^{(1)}_\mu$ given above
automatically satisfies Eq.~(\ref{Js-eq-n}) with $n=1$.
This means that Eq.~(\ref{Jmu-1-a}) is indeed the solution
{of  the first order under  global equilibrium conditions.}

Now let us turn to the second order solution that has not been considered before.
Similar to the way how we obtain the first order solution from the zeroth order,
the second order solution can be given by the iterative equation (\ref{Jmu-n}) with the
the first order solution (\ref{Jmu-1-a}),
\begin{eqnarray}
\label{Jmu-2}
\mathscr{J}_\mu^{(2)}&=& {J}_\mu^{(2)} \delta(p^2) + \frac{s}{2p^2}\epsilon_{\mu\nu\rho\sigma}p^\nu \nabla^\rho \mathscr{J}^{(1)\sigma}\nonumber\\
 &=& {X}_\mu^{(2)} \delta(p^2)
 +\frac{1}{4p^2}\left( p_\mu  \Omega_{\gamma\beta} p^\beta -p^2 \Omega_{\gamma\mu}\right)\Omega^{\gamma\lambda }p_\lambda f''\delta(p^2)\nonumber\\
& &{+}\frac{2}{p^6}\left( p_\mu   F_{\gamma\beta} p^\beta -p^2  F_{\gamma\mu}\right) F^{\gamma\lambda }p_\lambda f \delta(p^2)\nonumber\\
& &+\frac{1}{ p^4}\left( p_\mu  F_{\gamma\beta} p^\beta -p^2  F_{\gamma\mu}\right)\Omega^{\gamma\lambda }p_\lambda  f' \delta(p^2)\;.
\end{eqnarray}
Here ${X}_\mu^{(2)}$ can be constrained by inserting Eq.~(\ref{Jmu-2})
into  Eq.~(\ref{Js-c2-n}) with $n=2$. It turns out that
\begin{eqnarray}
\label{X-2}
\left(p_\mu {X}_\nu^{(2)}-p_\nu {X}_\mu^{(2)}\right)\delta(p^2) &=&0 \;,
\end{eqnarray}
which leads to $ {X}_\nu^{(2)}=0$, where we have used
 conditions (\ref{pd1}-\ref{pd2}) once again to arrive at the final result.
Now we finally obtain the second order solution
\begin{eqnarray}
\label{Jmu-2-a}
\mathscr{J}_\mu^{(2)}&=& - \frac{1}{4} \Omega_{\gamma\mu}\Omega^{\gamma\lambda }p_\lambda  f''\delta(p^2)
  - \frac{1}{4} p_\mu  \Omega_{\gamma\beta} p^\beta \Omega^{\gamma\lambda }p_\lambda f''\delta'(p^2)\nonumber\\
& &+  F_{\gamma\mu}\Omega^{\gamma\lambda }p_\lambda  f' \delta'(p^2)
+\frac{1}{2} p_\mu  F_{\gamma\beta} p^\beta \Omega^{\gamma\lambda }p_\lambda  f' \delta''(p^2)\nonumber\\
& &-  F_{\gamma\mu} F^{\gamma\lambda} p_\lambda f \delta''(p^2)
-\frac{1}{3} p_\mu   F_{\gamma\beta} p^\beta  F^{\gamma\lambda }p_\lambda f \delta'''(p^2) \;,
\end{eqnarray}
where we have used the identity
\begin{eqnarray}
p^6 \delta'''\left(p^2\right)&=&-3p^4\delta''\left(p^2\right)=6p^2 \delta'\left(p^2\right)=-6\delta(p^2) \;.
\end{eqnarray}

\section{Vector and axial currents}
\label{sec:current}

Once we have the Wigner function in phase space, the right-handed or left-handed current can be obtained directly by integrating
corresponding components of the Wigner function over the four-momentum
\begin{eqnarray}
j^\mu_s=\int d^4 p \mathscr{J}^\mu_s\;,
\end{eqnarray}
where we have recovered the chirality index $s$. The vector and axial currents are given by
\begin{eqnarray}
j^\mu= j^\mu_{+1} + j^\mu_{-1}\;,\ \ j^\mu_5= j^\mu_{+1} - j^\mu_{-1}\;.
\end{eqnarray}
Note that the vector current can also be called the fermion number or charge current,
while the axial current can also be called the chiral charge or chiral current.
The results for the zeroth and first order current are well-known
\begin{eqnarray}
\label{js-0-a}
j^{(0)\mu}_s &=& n_s u^\mu \;,\\
\label{js-1-a}
j^{(1)\mu}_s
&=&\xi_s \omega^\mu +\xi_{Bs} B^\mu \;,
\end{eqnarray}
where $n_s$ is the fermion number density, and $\xi_s$ and $\xi_{Bs}$ are transport coefficients associated with CVE and CME respectively in the right-handed and
 left-handed current $j^{\mu}_s$. They are given by
\begin{eqnarray}
\label{ns}
n_s &=& \frac{\mu_s}{6\pi^2}\left(\pi^2T^2+\mu^2_s\frac{}{}\right) \;,\\
\label{xis}
\xi_s &=& \frac{s }{12\pi^2}\left(\pi^2T^2+3\mu_s^2\frac{}{}\right) \;,\\
\label{xibs}
\xi_{Bs} &=& \frac{s }{4\pi^2}\mu_s \;.
\end{eqnarray}
In the zeroth order result $n_s$, we have dropped the infinite vacuum contribution.
In Eq.~(\ref{js-0-a}) the vorticity vector $\omega^\mu$ and the magnetic field vector $B^\mu$
are defined from the decomposition
\begin{eqnarray}
\label{FEB1}
F_{\mu\nu}&=& E_\mu u_\nu -E_\nu u_\mu + \epsilon_{\mu\nu\rho\sigma}u^\rho B^\sigma \;,\\
\label{VEB1}
T\Omega_{\mu\nu}&=& \varepsilon_\mu u_\nu -\varepsilon_\nu u_\mu
+ \epsilon_{\mu\nu\rho\sigma}u^\rho \omega^\sigma \;,
\end{eqnarray}
with
\begin{eqnarray}
E^\mu &=& F^{\mu\nu}u_\nu \;,\ \ \ \ \ B^\mu =  \tilde F^{\mu\nu}u_\nu =\frac{1}{2}\epsilon^{\mu\nu\alpha\beta}u_\nu F_{\alpha\beta}\;,\\
\varepsilon^\mu &=& T\Omega^{\mu\nu}u_\nu \;,\ \ \ \omega^\mu = T \tilde\Omega^{\mu\nu}u_\nu
=\frac{1}{2}\epsilon^{\mu\nu\alpha\beta}u_\nu \partial_\alpha^x u_\beta \;.
\end{eqnarray}
Similar to the electric or magnetic component of $F_{\mu\nu}$,
it is convenient to name $\varepsilon_\mu$ and $\omega_\mu$ as
the electric vorticity  and the magnetic vorticity, respectively.
It follows that the vector and axial current are given by
\begin{eqnarray}
\label{j-0-a}
j^{(0)\mu} &=& n u^\mu \;,\\
\label{j-1-a}
j^{(1)\mu} &=&  \xi \omega^\mu +\xi_B B^\mu \;,\\
\label{j5-0-a}
j^{(0)\mu}_5 &=& n_5 u^\mu \;,\\
\label{j5-1-a}
j^{(1)\mu}_5 &=&  \xi_5 \omega^\mu +\xi_{B5} B^\mu \;,
\end{eqnarray}
with
\begin{eqnarray}
n &=&\frac{\mu}{3\pi^2}\left(\pi^2T^2 +\mu^2 +3 \mu_5^2\frac{}{}\right)\;,\ \
n_5 =\frac{\mu_5}{3\pi^2}\left(\pi^2T^2 +3\mu^2 + \mu_5^2\frac{}{}\right)\;,\nonumber\\
\xi &=& \frac{\mu\mu_5 }{\pi^2}\;,\ \ \ \xi_B = \frac{\mu_5}{2\pi^2}\;,\ \ \
\xi_5 = \frac{1}{6\pi^2}\left[\pi^2T^2+3(\mu^2 + \mu_5^2)\frac{}{}\right]\;,\ \ \
\xi_{B5} = \frac{\mu}{2\pi^2} \;.
\label{n-n5-xi-xi5}
\end{eqnarray}
where $n$ and $n_5$ are the fermion number (charge) and chiral charge density respectively,
and $\xi$, $\xi_B$, $\xi_5$ and $\xi_{B5}$ are well-known anomalous transport coefficients associated with CVE, CME, LPE and CSE, respectively.

The second-order current can be obtained by integrating Eq.~(\ref{Jmu-2-a})
over the four-momentum,
\begin{eqnarray}
\label{j-2}
j^{(2)\mu}_{s}
&=&-\frac{1}{4}\Omega^{\gamma\mu}\Omega_{\gamma\lambda } u^\lambda  \int d^4 p (u\cdot p) f^{\prime\prime}_s\delta(p^2)
-\frac{1}{4}u^\mu u^\beta u_\lambda
\Omega_{\gamma\beta}\Omega^{\gamma\lambda } \int d^4 p  (u\cdot p)^3  f''_s\delta'(p^2) \nonumber\\
& &-  \frac{1}{12} \left(\Delta^{\mu\beta} u_\lambda +{ \Delta^{\mu}}_\lambda u^\beta+{\Delta_{\lambda}}^\beta u^\mu\right) \Omega_{\gamma\beta}\Omega^{\gamma\lambda }
\int d^4p  (u\cdot p)\bar p^2  f''_s\delta'(p^2)\nonumber\\
& & + F^{\gamma\mu}\Omega_{\gamma\lambda } u^\lambda  \int d^4 p  (u\cdot p)  f'_s\delta'(p^2)
+\frac{1}{2} u^\mu u^\beta u_\lambda  F_{\gamma\beta}\Omega^{\gamma\lambda } \int d^4p  (u\cdot p)^3  f'_s\delta''(p^2)  \nonumber\\
& &+  \frac{1}{6} \left(\Delta^{\mu\beta} u_\lambda + {\Delta^\mu}_\lambda u^\beta+{\Delta_{\lambda}}^\beta u^\mu\right)F_{\gamma\beta} \Omega^{\gamma\lambda }
\int d^4p (u\cdot p)\bar p^2  f'_s\delta''(p^2)\nonumber\\
& & - F^{\gamma\mu} F_{\gamma\lambda } u^\lambda  \int d^4p  (u\cdot p)   f_s \delta''(p^2)
-\frac{1}{3} u^\mu u^\beta u_\lambda  F_{\gamma\beta} F^{\gamma\lambda } \int d^4p (u\cdot p)^3  f_s \delta'''(p^2)   \nonumber\\
& & - \frac{1}{9}  \left({\Delta}^{\mu\beta} u_\lambda + {\Delta^\mu}_\lambda u^\beta+{\Delta_{\lambda}}^\beta u^\mu\right) F_{\gamma\beta} F^{\gamma\lambda } \int d^4p  (u\cdot p)\bar p^2 f_s \delta'''(p^2) \;.
\end{eqnarray}
In the above equation we have used following moment identities
\begin{eqnarray}
\int d^4 p\,  p^\lambda Y &=& u^\lambda \int d^4 p\,  (u\cdot p) Y \;,\\
\int d^4 p\, p^\mu p^\beta p^\lambda Y &=& u^\mu u^\beta u^\lambda  \int d^4 p\,  (u\cdot p)^3 Y \nonumber\\
& &+\frac{1}{3}\left(\Delta^{\mu\beta}u^\lambda + \Delta^{\mu\lambda}u^\beta +\Delta^{\lambda\beta}u^\mu\right) \int d^4 p\,  (u\cdot p)\bar p^2 Y,
\end{eqnarray}
where $Y$ can be any scalar functions of $u\cdot p$ and $p^2$,
$\Delta^{\mu\nu}=g^{\mu\nu}-u^\mu u^\nu$ and $\bar p^\mu  = \Delta^{\mu\nu}p_\nu$.
Using the decomposition (\ref{FEB1}) and (\ref{VEB1}), $j^{(2)}_{s,\mu}$ can be put into the form
\begin{eqnarray}
j^{(2)\mu}_{s}
&=&u^\mu\left(\varepsilon ^2 +\omega^2\right)\frac{1}{6} \int d^4p  (u\cdot p)\bar p^2 f''_s\delta'(p^2) \nonumber\\
& &-  \epsilon^{\gamma\mu\rho\sigma}  \varepsilon_\rho u_\sigma \omega_\gamma
\int d^4 p (u\cdot p)f''_s\left[ \frac{ 1 }{ 6} \bar p^2 \delta'(p^2) + \frac{1}{4}\delta(p^2)\right]\nonumber\\
& &-u^\mu ( \varepsilon \cdot E + \omega\cdot B )\frac{1}{3}\int d^4 p (u\cdot p)\bar p^2  f'_s\delta''(p^2) \nonumber\\
& &+\epsilon^{\gamma\mu\rho\sigma} E_\rho u_\sigma\omega_\gamma
\int d^4 p (u\cdot p)f'_s \left[ \frac{1}{3} \bar p^2 \delta''(p^2) + \delta'(p^2) \right] \nonumber\\
& & +u^\mu\left( E^2 + B^2 \right)\frac{2}{9}\int d^4 p  (u\cdot p)\bar p^2   f_s \delta'''(p^2) \nonumber\\
& & -  \epsilon^{\gamma\mu\rho\sigma} u_\rho B_\sigma  E_\gamma
\int d^4 p (u\cdot p) f_s \left[\frac{2}{9} \bar p^2  \delta'''(p^2) + \delta''(p^2) \right]\;.
\label{j2-int}
\end{eqnarray}
After completing integrals in Eq.~(\ref{j2-int}), we obtain  second-order currents
\begin{eqnarray}
j^{(2)\mu}_s &=&-\frac{\mu_s }{4\pi^2} ( \varepsilon^2+\omega^2 )u^\mu
-\frac{1}{8\pi^2} ( \varepsilon\cdot E + \omega\cdot B ) u^\mu
- \frac{ C_s}{24\pi^2} ( E^2+B^2 )u^\mu \nonumber\\
& &-\frac{1}{8\pi^2} \epsilon^{\mu \nu \rho\sigma} u_\nu E_\rho \omega_\sigma
-\frac{ C_s}{12\pi^2} \epsilon^{\mu\nu\rho\sigma} u_\nu  E_\rho B_\sigma \;,
\label{j-2-a}
\end{eqnarray}
where
\begin{eqnarray}
\label{Cs}
C_s &=&\frac{1}{T}\int_0^\infty \frac{d p_0}{p_0}  \left[\frac{e^{p_0/T-\bar\mu_s}}{\left(e^{p_0/T-\bar\mu_s}+1\right)^2}
-\frac{e^{p_0/T+\bar\mu_s}}{\left(e^{p_0/T+\bar\mu_s}+1\right)^2}\right] \;.
\end{eqnarray}
It is obvious that $C_s$ is an odd function of $\bar\mu_s$.
When $|\bar\mu_s|\ll 1$ or at high temperature limit,
we can expand the integrand in Eq.~(\ref{Cs}) in serials and work out the integral analytically
\begin{equation}
C_s = -\frac{14\zeta'(-2)\mu_s}{T^2}\approx \frac{0.4263\mu_s}{ T^2} \;.
\end{equation}
When $|\bar\mu_s|\gg 1$ or at low temperature limit, we can approximate the Fermi-Dirac distribution function
by a step function $\Theta(\pm\mu_s-p_0)$ and obtain the analytic result
\begin{equation}
C_s = \frac{1}{\mu_s} \;.
\end{equation}
From Eq.~(\ref{j-2-a}) the vector and axial current are given by
\begin{eqnarray}
\label{jv-2}
j^{(2)\mu}
&=&-\frac{\mu}{2\pi^2} ( \varepsilon^2+\omega^2 )u^\mu
-\frac{1}{4\pi^2} ( \varepsilon\cdot E + \omega\cdot B ) u^\mu
-\frac{C}{12\pi^2}  ( E^2+B^2 )u^\mu\nonumber\\
& &-\frac{1}{4\pi^2} \epsilon^{\mu \nu \rho\sigma} u_\nu E_\rho \omega_\sigma
-\frac{C}{6\pi^2 } \epsilon^{\mu\nu\rho\sigma} u_\nu  E_\rho B_\sigma \;,\\
\label{j5-2}
j^{(2)\mu}_5
&=&-\frac{\mu_5}{2\pi^2} ( \varepsilon^2+\omega^2 )u^\mu
-\frac{C_5}{12\pi^2}  ( E^2+B^2 )u^\mu
-\frac{C_5}{6\pi^2 } \epsilon^{\mu\nu\rho\sigma} u_\nu  E_\rho B_\sigma \;,
\end{eqnarray}
with
\begin{equation}
C = \frac{1}{2}\left(C_{+1}+C_{-1}\right)\;, \ \ \ C_5 =\frac{1}{2}\left(C_{+1}-C_{-1}\right)\;.
\end{equation}
When $|\mu_s|\ll T$, we have
\begin{eqnarray}
C = -\frac{14\zeta'(-2)\mu}{T^2}\approx \frac{0.4263\mu}{T^2}\;,\ \
C_5 = -\frac{14\zeta'(-2)\mu_5}{T^2}\approx \frac{0.4263\mu_5}{ T^2} \;.
\end{eqnarray}
When $|\mu_s|\gg T$, we have
\begin{eqnarray}
C = \frac{\mu}{ (\mu^2-\mu_5^2)}\;,\ \ C_5 = - \frac{\mu_5}{(\mu^2-\mu_5^2)} \;.
\end{eqnarray}

Now we look closely at the vector current (\ref{jv-2}).
The first line of (\ref{jv-2}) indicates that the charge density is modified by quadratic terms
$\varepsilon^2$, $\omega^2$, $ E^2 $, $B^2$, $ \varepsilon\cdot E$ and $ \omega\cdot B$.
The second line of (\ref{jv-2}) are the Hall currents induced along the direction orthogonal
to both $E^\mu$ and $\omega^\nu$ or that orthogonal to both $E^\mu$ and $B^\nu$
in the comoving frame of the fluid cell.
It is interesting to observe that there is no Hall current induced by $\varepsilon^\mu$ and $\omega^\nu$.
It should be clarified here that the mixed Hall current
$\epsilon^{\mu \nu \rho\sigma} u_\nu E_\rho \omega_\sigma$ is actually identical to
$\epsilon^{\mu \nu \rho\sigma} u_\nu \varepsilon_\rho B_\sigma$
in this paper due to the integrability  condition (\ref{integrability}) which is equivalent to
\begin{eqnarray}
\label{integrability-1}
\epsilon_{\mu\nu\rho\sigma}\left(E^\rho\omega^\sigma - \varepsilon^\rho B^\sigma\right) &=& 0\ \ \textrm{ and }\ \ \
\epsilon_{\mu\nu\rho\sigma}\left(E^\rho \varepsilon^\sigma + \omega^\rho B^\sigma \right) = 0 \;.
\end{eqnarray}

For the axial current, the first and second terms in (\ref{j5-2}) indicate that the axial charge density gets modified
by quadratic terms $\varepsilon^2$, $\omega^2$, $ E^2 $ and $B^2$ but not from mixed terms
$\varepsilon\cdot E$ and $ \omega\cdot B$ due to the symmetry which is different from the charge density.
The last term in (\ref{j5-2}) is the axial Hall current generated by $E^\mu$ and $B^\nu$ only, but
not by $E^\mu$ and $\omega^\nu$ or $\varepsilon^\mu$ and $B^\nu$, which is also different from the vector current.
Like the charge current, there is no axial Hall current from $\varepsilon^\mu$ and $\omega^\nu$.

\section{Energy-momentum tensor}
\label{sec:tensor}
In the Wigner function formalism, the stress or energy-momentum tensor can be from the vector component as
\begin{eqnarray}
T^{\mu\nu}=\int d^4 p \mathscr{V}^\mu p^\nu
=\int d^4 p \left(\mathscr{J}^\mu_{+1}+\mathscr{J}^\mu_{-1}\right) p^\nu \;.
\end{eqnarray}
Note that this is the canonical definition  and is not necessarily symmetric.
The results for  the stress tensor of the right-handed or left-handed part
at the zeroth  and first order are,
\begin{eqnarray}
T^{(0)\mu\nu}_s&=&\int d^4 p \mathscr{J}_{s}^{(0)\mu} p^\nu = u^\mu u^\nu \rho_s -\frac{1}{3}\Delta^{\mu\nu}\rho_s \;,\\
T_{s}^{(1)\mu\nu}&=&\int d^4 p \mathscr{J}_{s}^{(1)\mu} p^\nu \nonumber\\
&=&  s n_s \left(u^\mu \omega^\nu + u^\nu \omega^\mu \right)
+\frac{\xi_s}{2}\left(u^\mu B^\nu + u^\nu B^\mu -\epsilon^{\mu\nu\alpha\beta}u_\alpha E_\beta \right)  \nonumber\\
&&-\frac{ s n_s}{2}\left(u^\mu \omega^\nu - u^\nu \omega^\mu +\epsilon^{\mu\nu\alpha\beta}u_\alpha
\varepsilon_\beta \right) \;,
\end{eqnarray}
where $n_s$ and $\xi_s$ in $T^{(1)\mu\nu}_s$ are given by Eqs.~(\ref{ns},\ref{xis}),
and the energy density $\rho _s$ in $T^{(0)\mu\nu}_s$ is
\begin{eqnarray}
\rho_s &=&\frac{ T^4}{2\pi^2}\left(\frac{7}{60}\pi^4+\frac{1}{2}\pi^2 \bar\mu_s^2+\frac{1}{4}{\bar\mu_s^{4}}\right)\;,
\end{eqnarray}
where we have dropped the infinite vacuum energy density.
After taking a sum of the right-handed and left-handed contributions, we obtain the total energy-momentum tensor
\begin{eqnarray}
\label{T0}
T^{(0)\mu\nu}&=&T^{(0)\mu\nu}_{+1} + T^{(0)\mu\nu}_{-1} = \rho  u^\mu u^\nu -\frac{1}{3}\rho\Delta^{\mu\nu} ,\\
\label{T1}
T^{(1)\mu\nu}&=&T^{(0)\mu\nu}_{+1} + T^{(0)\mu\nu}_{-1} \nonumber\\
&=& n_5 \left(u^\mu \omega^\nu + u^\nu \omega^\mu \right)
+\frac{\xi}{2}\left(u^\mu B^\nu + u^\nu B^\mu -\epsilon^{\mu\nu\alpha\beta}u_\alpha E_\beta  \right)  \nonumber\\
& &-\frac{ n_5}{2}\left(u^\mu \omega^\nu - u^\nu \omega^\mu +\epsilon^{\mu\nu\alpha\beta}u_\alpha \varepsilon_\beta \right) \;,
\end{eqnarray}
where $n$ and $\xi$ in $T^{(1)\mu\nu}$ are given by Eq.~(\ref{n-n5-xi-xi5}),
and the energy density in $T^{(0)\mu\nu}$ is
\begin{eqnarray}
\rho &=&\frac{ T^4}{4\pi^2}\left[\frac{7}{15}\pi^4+ 2 \pi^2 (\bar\mu^2+\bar\mu_5^2)
+\bar \mu^4 + 6\bar \mu^2 \bar \mu_5^2+\bar\mu_5^4 \right] \;.
\end{eqnarray}

Now let us compute the stress tensor at the second order. We decompose the stress tensor into three parts,
\begin{eqnarray}
T_{s}^{(2)\mu\nu}=T_{s,\textrm{vv}}^{(2)\mu\nu}+T_{s,\textrm{ve}}^{(2)\mu\nu}+T_{s,\textrm{ee}}^{(2)\mu\nu}\;,
\end{eqnarray}
which `v' means the vorticity and `e' means the electromagnetic field,
so these three terms are coupling terms of the vorticity-vorticity, vorticity-electromagnetic-field and
electromagnetic-field-electromagnetic-field, respectively. These terms are given by
\begin{eqnarray}
T_{s,\textrm{vv}}^{(2)\mu\nu}&=&-\frac{1}{4}{\Omega^{\gamma}}_\beta \Omega_{\gamma\lambda }
\int d^4 p\ p^\mu  p^\nu p^\beta p^\lambda f'' \delta'(p^2)
-\frac{1}{4}  \Omega^{\gamma\mu}  \Omega_{\gamma\lambda }\int d^4 p\  p^\nu p^\lambda f'' \delta(p^2)\;,\\
T_{s,\textrm{ve}}^{(2)\mu\nu}&=&\frac{1}{2}{F^{\gamma}}_{\beta} \Omega_{\gamma\lambda }
\int d^4 p\ p^\mu  p^\nu p^\beta p^\lambda f' \delta''(p^2)
+ F^{\gamma\mu}  \Omega_{\gamma\lambda }\int d^4 p\  p^\nu p^\lambda f' \delta'(p^2)\;,\\
\label{T-s-ee}
T_{s,\textrm{ee}}^{(2)\mu\nu}&=&-\frac{1}{3}  {F^{\gamma}}_\beta F_{\gamma\lambda } \int d^4 p\ p^\mu  p^\nu p^\beta p^\lambda f \delta'''(p^2)
- F^{\gamma\mu} F_{\gamma\lambda }\int d^4 p\  p^\nu p^\lambda f \delta''(p^2)\;.
\end{eqnarray}
Using  moment identities
\begin{eqnarray}
\int d^4 p\, p^\nu  p^\lambda Y &=& u^\nu u^\lambda \int d^4 p\,  (u\cdot p)^2 Y +\frac{1}{3}\Delta^{\mu\nu}\int d^4 p\, \bar p^2 Y,\\
\int d^4 p\, p^\mu p^\nu p^\beta p_\lambda Y &=& u^\mu u^\nu u^\beta u^\lambda  \int d^4 p\,  (u\cdot p)^4 Y \nonumber\\
& &+\frac{1}{15}\left(\Delta^{\mu\nu} \Delta^{\beta\lambda}+ \Delta^{\mu\beta} \Delta^{\nu\lambda}+\Delta^{\mu\lambda}  \Delta^{\beta\nu}\right)
 \int d^4 p\, \bar p^4 Y\nonumber\\
& &+\frac{1}{3}\left(u^\mu u^\nu \Delta^{\beta\lambda}+u^\beta u^\lambda \Delta^{\mu\nu}
+u^\mu u^\beta \Delta^{\nu\lambda}+u^\mu u^\lambda \Delta^{\beta\nu} \right.\nonumber\\
& &\left.+u^\nu u^\lambda \Delta^{\beta\mu}+u^\nu u^\beta \Delta^{\mu\lambda}\right)
\int d^4 p (u\cdot p)^2\bar p^2 Y \;,
\end{eqnarray}
and the decomposition (\ref{FEB1}) and (\ref{VEB1}), we can write  the first two equations as

\begin{eqnarray}
T_{s,\textrm{vv}}^{(2)\mu\nu}&=&-\frac{\beta^2}{4}u^\mu u^\nu \varepsilon^2
\int d^4 p\ (u\cdot p)^4 f'' \delta'(p^2)\nonumber\\
& &-\frac{\beta^2}{60}\left[\Delta^{\mu\nu} (\varepsilon^2 - 4\omega^2)
+ 2\varepsilon^\mu \varepsilon^\nu
+ 2\omega^\mu \omega^\nu \right]  \int d^4 p\ \bar p^4 f'' \delta'(p^2)\nonumber\\
& &-\frac{\beta^2}{12}\left[u^\mu u^\nu(\varepsilon^2 - 2\omega^2)
+\Delta^{\mu\nu}\varepsilon^2 +
2( u^\mu \epsilon^{\nu\alpha\beta\gamma} +u^\nu \epsilon^{\mu\alpha\beta\gamma} )
 u_\alpha \varepsilon_\beta \omega_\gamma \right]\nonumber\\
& &\hspace{1cm}\int d^4 p\ (u\cdot p)^2 \bar p^2 f'' \delta'(p^2)\nonumber\\
& & -\frac{\beta^2}{4}\left(u^\mu u^\nu \varepsilon^2 + u^\nu \epsilon^{\mu\alpha\beta\gamma}u_\alpha \varepsilon_\beta \omega_\gamma\right)  \int d^4 p\  (u\cdot p)^2 f'' \delta(p^2)\nonumber\\
& & -\frac{\beta^2}{12} \left( \varepsilon^\mu \varepsilon^\nu
+ \omega^\mu \omega^\nu -\Delta^{\mu\nu} \omega^2
 + u^\mu\epsilon^{\nu\alpha\beta\gamma} u_\alpha \varepsilon_\beta \omega_\gamma  \right)
 \int d^4 p\  \bar p^2 f'' \delta(p^2),
\end{eqnarray}
\begin{eqnarray}
T_{s,\textrm{ve}}^{(2)\mu\nu}&=&\beta u^\mu u^\nu \varepsilon\cdot E
\int d^4 p\ (u\cdot p)^4 f' \delta''(p^2)\nonumber\\
& &+\frac{\beta}{30}\left[\Delta^{\mu\nu} (\varepsilon\cdot E - 4\omega\cdot B)
 +2 E^\mu \varepsilon^\nu
+ 2 \omega^\mu B^\nu \right]  \int d^4 p\ \bar p^4 f' \delta''(p^2)\nonumber\\
& &+\frac{\beta}{3}\left[u^\mu u^\nu(\varepsilon\cdot E - 2\omega\cdot B)
+\Delta^{\mu\nu}\varepsilon\cdot E +
2( u^\mu \epsilon^{\nu\alpha\beta\gamma} +u^\nu \epsilon^{\mu\alpha\beta\gamma} )
 u_\alpha E_\beta \omega_\gamma \right]\nonumber\\
& &\hspace{1cm} \int d^4 p\ (u\cdot p)^2 \bar p^2 f' \delta''(p^2)\nonumber\\
& & +\beta \left(u^\mu u^\nu \varepsilon\cdot E + u^\nu \epsilon^{\mu\alpha\beta\gamma}u_\alpha E_\beta \omega_\gamma\right)
\int d^4 p\  (u\cdot p)^2 f' \delta'(p^2)\nonumber\\
& & +\frac{\beta}{3} \left( E^\mu \varepsilon^\nu
+ \omega^\mu B^\nu -\Delta^{\mu\nu} \omega\cdot B
 + u^\mu\epsilon^{\nu\alpha\beta\gamma} u_\alpha E_\beta \omega_\gamma  \right)
 \int d^4 p\  \bar p^2 f' \delta'(p^2).
\end{eqnarray}
Completing integration and collecting  similar terms,  we  obtain
\begin{eqnarray}
T_{s,\textrm{vv}}^{(2)\mu\nu}&=&-\frac{s}{2}\xi_s
\left[{3 u^\mu u^\nu ( \omega^2+\varepsilon^2) -\Delta^{\mu\nu} ( \omega^2+\varepsilon^2)}
 -2( u^\mu \epsilon^{\nu\alpha\beta\gamma} + u^\nu \epsilon^{\mu\alpha\beta\gamma} ) u_\alpha \varepsilon_\beta \omega_\gamma\right. \nonumber\\
& & \left. -2 (u^\mu \epsilon^{\nu\alpha\beta\gamma} - u^\nu \epsilon^{\mu\alpha\beta\gamma})  u_\alpha \varepsilon_\beta \omega_\gamma \right]\;,\\
T_{s,\textrm{ve}}^{(2)\mu\nu}
&=&-\frac{s}{2}\xi_{Bs} \left[ u^\mu u^\nu (\omega\cdot B+\varepsilon\cdot E)
- (\omega^\mu B^\nu + E^\mu \varepsilon^\nu) \right. \nonumber\\
&&\left. -(u^\mu \epsilon^{\nu\alpha\beta\gamma}+u^\nu \epsilon^{\mu\alpha\beta\gamma})u_\alpha E_\beta \omega_\gamma
 - 2 (u^\mu \epsilon^{\nu\alpha\beta\gamma}-u^\nu \epsilon^{\mu\alpha\beta\gamma})u_\alpha E_\beta \omega_\gamma \right]\;,
\label{ts-ve}
\end{eqnarray}
However, when we deal with $T_{s,\textrm{ee}}^{(2)\mu\nu}$, we find that it has
logarithmic ultraviolet divergence which has to be regularized and renormalized.  The regularization with a naive
momentum cutoff will break Lorentz invariance  and destroy the energy momentum conservation. To avoid such a problem,
we apply  dimensional regularization. Now let us make a tensor decomposition
in $d=4-\epsilon$ dimension with a small positive number $\epsilon$.
\begin{eqnarray}
\int d^d p\, p^\nu  p^\lambda Y &=& u^\nu u^\lambda \int d^d p\,  (u\cdot p)^2 Y +\frac{1}{d-1}\Delta^{\mu\nu}\int d^d p\, \bar p^2 Y,\\
\int d^d p\, p^\mu p^\nu p^\beta p^\lambda Y &=& u^\mu u^\nu u^\beta u^\lambda  \int d^d p\,  (u\cdot p)^4 Y \nonumber\\
& &+\frac{1}{d^2-1}\left(\Delta^{\mu\nu} \Delta^{\beta\lambda}+ \Delta^{\mu\beta} \Delta^{\nu\lambda}+\Delta^{\mu\lambda}  \Delta^{\beta\nu}\right) \int d^d p\, \bar p^4 Y\nonumber\\
& &+\frac{1}{d-1}\left(u^\mu u^\nu \Delta^{\beta\lambda}+u^\beta u^\lambda \Delta^{\mu\nu}
+u^\mu u^\beta \Delta^{\nu\lambda}+u^\mu u^\lambda \Delta^{\beta\nu} \right.\nonumber\\
& &\left.+u^\nu u^\lambda \Delta^{\beta\mu}+u^\nu u^\beta \Delta^{\mu\lambda}\right)
\int d^d p (u\cdot p)^2\bar p^2 Y \;,
\end{eqnarray}
With such tensor decomposition in $d=4-\epsilon$ dimension,  we can write Eq.~(\ref{T-s-ee}) as
\begin{eqnarray}
\label{T-s-ee-1}
T_{s,{\textrm{ee}}}^{(2)\mu\nu}
&=&-\frac{1}{3} u^\mu u^\nu E^2 \int d^{4-\epsilon} p\ (u\cdot p)^4 f \delta'''(p^2)\nonumber\\
& &-\frac{1}{45-24\epsilon} \left( \Delta^{\mu\nu} {F_{\gamma}}^\beta F^{\gamma\lambda }\Delta_{\beta\lambda}
+2 \Delta^{\mu\kappa}\Delta^{\nu\lambda}{F_{\gamma\kappa}} {F^{\gamma}}_\lambda  \right)\int d^{4-\epsilon} p\ \bar p^4 f \delta'''(p^2) \nonumber\\
& &-\frac{1}{9-3\epsilon}\left( u^\mu u^\nu {F_{\gamma}}^\beta {F^{\gamma\lambda }}\Delta_{\beta \lambda}
+ \Delta^{\mu\nu} E^2 + 2u^\mu \Delta^{\nu\lambda}E^\gamma F_{\gamma\lambda}
+2 u^\nu \Delta^{\mu\lambda}E^\gamma F_{\gamma\lambda} \right)\nonumber\\
& &\hspace{1.2cm}\times\int d^{4-\epsilon} p\ (u\cdot p)^2 \bar p^2 f \delta'''(p^2)\nonumber\\
& &-  u^\nu F^{\gamma\mu} E_{\gamma}\int d^4 p\  (u\cdot p)^2 f \delta''(p^2)
-\frac{1}{3-\epsilon}\Delta^{\nu\lambda } F^{\gamma\mu} F_{\gamma\lambda}\int d^{4-\epsilon} p\  \bar p^2 f \delta''(p^2)
\end{eqnarray}
We give  details  of  the integration in the first term as an example in Appendix \ref{sec:Integation}.
After integrating over momentum, we  obtain  the pure electromagnetic part
of the energy-momentum tensor
\begin{eqnarray}
\label{ts-ee}
T_{s,\textrm{ee}}^{(2)\mu\nu}&=&
-\frac{1}{12}\kappa_s^\epsilon\left(  \frac{1}{4}g^{\mu\nu} {F_{\gamma\beta}} {F^{\gamma\beta }} - F^{\gamma\mu}{F_\gamma}^\nu\right)
+ \frac{1}{48\pi^2} u^\mu u^\nu E^2
-\frac{1}{48\pi^2}\Delta^{\mu\nu} \left(E^2+2B^2\right)\nonumber\\
& &+\frac{1}{12\pi^2}\left( E^\mu E^\nu +B^\mu B^\nu \right)
+\frac{1}{16\pi^2}\left( u^\mu \epsilon^{\nu\alpha\beta\gamma}
+ u^\nu \epsilon^{\mu\alpha\beta\gamma}   \right)u_\alpha E_\beta B_\gamma\nonumber\\
& &+\frac{1}{16\pi^2}\left(u^\mu \epsilon^{\nu\alpha\beta\gamma}
- u^\nu \epsilon^{\mu\alpha\beta\gamma}   \right)u_\alpha E_\beta B_\gamma,
\end{eqnarray}
where $\kappa_s^\epsilon$ is given by
\begin{eqnarray}
\label{kappa-s-epsilon}
\kappa_s^\epsilon
&=&\frac{4\pi^{(3-\epsilon)/{2}} T^{-\epsilon}}{\Gamma\left((3-\epsilon)/{2}\right)(2\pi)^{3-\epsilon}}
\int_0^\infty  \frac{dy}{y^{1+\epsilon}}\left[\frac{1}{e^{(y-\bar\mu_s)}+1}+\frac{1}{e^{(y+\bar\mu_s)}+1}{-1}\right].
\end{eqnarray}
We can expand $\kappa_s^\epsilon$ around $\epsilon=0$ as
\begin{eqnarray}
\label{kappa}
\kappa_s^\epsilon&=&-\frac{1}{\pi^2 }\left[\frac{1}{\epsilon}+
\ln 2  +\frac{1}{2} \ln \pi + \frac{1}{2}\psi\left(\frac{3}{2}\right) - \ln T + \hat\kappa_s\right],
\end{eqnarray}
where $\psi(x)$ is the digamma  function and $\hat\kappa_s$ is given by
\begin{eqnarray}
\label{hatkappa}
\hat \kappa_s &=&
\int_0^\infty  dy \,\ln y\ \frac{d}{dy}\left[\frac{1}{e^{(y-\bar\mu_s)}+1}+\frac{1}{e^{(y+\bar\mu_s)}+1}\right].
\end{eqnarray}
Obviously, we can see in (\ref{hatkappa}) that the integral in (\ref{kappa-s-epsilon}) contains logarithmic ultraviolet
divergence at the limit $\epsilon\rightarrow 0$.
The coefficient $\kappa^\epsilon_s$ or $\hat \kappa_s$ is an even function of $\bar\mu_s$.
It is also easy to verify
\begin{eqnarray}
\left.\epsilon\kappa^\epsilon_s\right|_{\epsilon\rightarrow 0} = -\frac{1}{\pi^2} ,\ \ \
\left. \frac{d \kappa^\epsilon_s(\bar\mu_s)}{d \beta}\right|_{\epsilon\rightarrow 0} = -\frac{T}{\pi^2},\ \ \
 \left. \frac{d \kappa^\epsilon_s(\bar\mu_s)}{d\bar \mu_s}\right|_{\epsilon\rightarrow 0} =  T C_s(\bar\mu_s) \;.
\end{eqnarray}
Taking a sum over  contributions from  left-handed and right-handed fermions, the total stress tensor is given by
\begin{eqnarray}
\label{tvv2-munu}
T_{\textrm{vv}}^{(2)\mu\nu}&=&-\frac{1}{2}\xi_5
\left[{3 u^\mu u^\nu ( \omega^2+\varepsilon^2)  -\Delta^{\mu\nu} ( \omega^2+\varepsilon^2)}
 -2( u^\mu \epsilon^{\nu\alpha\beta\gamma} + u^\nu \epsilon^{\mu\alpha\beta\gamma} ) u_\alpha \varepsilon_\beta \omega_\gamma\right. \nonumber\\
& &\left. -2 (u^\mu \epsilon^{\nu\alpha\beta\gamma} - u^\nu \epsilon^{\mu\alpha\beta\gamma})  u_\alpha \varepsilon_\beta \omega_\gamma \right],\\
\label{tve2-munu}
T_{\textrm{ve}}^{(2)\mu\nu}
&=&-\frac{1}{2}\xi_{B5} \left[u^\mu u^\nu (\omega\cdot B+\varepsilon\cdot E)
- (\omega^\mu B^\nu + E^\mu \varepsilon^\nu)
-(u^\mu \epsilon^{\nu\alpha\beta\gamma}+u^\nu \epsilon^{\mu\alpha\beta\gamma})u_\alpha E_\beta \omega_\gamma  \right.\nonumber\\
& &\left. - 2 ( u^\mu \epsilon^{\nu\alpha\beta\gamma}
-   u^\nu \epsilon^{\mu\alpha\beta\gamma} )u_\alpha E_\beta \omega_\gamma \right],\\
\label{tee2-munu-0}
T_{\textrm{ee}}^{(2)\mu\nu}
&=&-\frac{1}{6}\kappa^\epsilon \left( \frac{1}{4}g^{\mu\nu} {F_{\gamma\beta}} {F^{\gamma\beta } -  F^{\gamma\mu}{F_\gamma}^\nu }\right)
+ \frac{1}{24\pi^2}\left[ u^\mu u^\nu E^2
-\Delta^{\mu\nu} \left(E^2+2B^2\right)\right.\nonumber\\
& &+4\left( E^\mu E^\nu +B^\mu B^\nu \right)
+3\left( u^\mu \epsilon^{\nu\alpha\beta\gamma}
+ u^\nu \epsilon^{\mu\alpha\beta\gamma}   \right)u_\alpha E_\beta B_\gamma\nonumber\\
& &\left.+3\left(u^\mu \epsilon^{\nu\alpha\beta\gamma}
- u^\nu \epsilon^{\mu\alpha\beta\gamma}   \right)u_\alpha E_\beta B_\gamma \right]\;,
\end{eqnarray}
where $\kappa^\epsilon = (\kappa_{+1}^\epsilon  +  \kappa_{-1}^\epsilon)/2$.

We see that  energy density and  pressure are  modified in vorticity
 and electromagnetic fields at the second order, which are not the case at the first order.
Similar to the first order result  in Eq.(\ref{T1}), there are also antisymmetric contributions
to the energy-momentum tensor at the second order. It is straightforward
to verify with Eqs.(\ref{T0}-\ref{T1},\ref{tvv2-munu}-\ref{tee2-munu-0}) that
 the trace of the total energy-momentum
tensor up to the second order  vanishes
\begin{equation}
g_{\mu\nu}T^{\mu\nu}=0,
\end{equation}
free of trace anomaly. Note  that in taking the trace of $T_{\textrm{ee}}^{(2)\mu\nu}$ we used
$g_{\mu\nu}g^{\mu\nu}=4-\epsilon$. The trace anomaly does not arise here because
the electromagnetic field in our work is only a classical background field and
keeps scale invariance.
We  note that the divergent part in $T_{\textrm{ee}}^{(2)\mu\nu}$ is  proportional to the stress tensor
of the free electromagnetic field. It is remarkable that this  divergent term $\sim 1/\epsilon$  is exactly  the contribution from the quantum correction of the electromagnetic field \cite{Peskin:1995ev} but with a wrong sign. Hence we can add this contribution to our result,
cancel the divergent term and arrive at the renormalized finite result

 \begin{eqnarray}
 \label{tee2-munu-re}
T_{\textrm{ee,Re}}^{(2)\mu\nu}
&=&\frac{1}{6\pi^2}\left(\hat\kappa + \ln \frac{\Lambda}{T}\right)
\left( \frac{1}{4}g^{\mu\nu} {F_{\gamma\beta}} {F^{\gamma\beta } -  F^{\gamma\mu}{F_\gamma}^\nu }\right)
+ \frac{1}{24\pi^2}\left[ u^\mu u^\nu E^2
-\Delta^{\mu\nu} \left(E^2+2B^2\right)\right.\nonumber\\
& &+4\left( E^\mu E^\nu +B^\mu B^\nu \right)
+3\left( u^\mu \epsilon^{\nu\alpha\beta\gamma}
+ u^\nu \epsilon^{\mu\alpha\beta\gamma}   \right)u_\alpha E_\beta B_\gamma\nonumber\\
& &\left.+3\left(u^\mu \epsilon^{\nu\alpha\beta\gamma}
- u^\nu \epsilon^{\mu\alpha\beta\gamma}   \right)u_\alpha E_\beta B_\gamma \right]
\;,
\end{eqnarray}
where $\hat\kappa = \hat\kappa_{+1} +  \hat\kappa_{-1}$ and $\Lambda$ is a renormalization scale
in quantum  electromagnetic field. Note that we have also absorbed all possible remaining constant terms in Eq.~(\ref{kappa})
into $\Lambda$. After removing the divergence in $T_{\textrm{ee}}^{(2)\mu\nu}$,
we can safely calculate the trace of the energy-momentum tensor in four dimensions and obtain
the trace anomaly
\begin{equation}
g_{\mu\nu}T_{\textrm{ee,Re}}^{(2)\mu\nu}=\frac{1}{24\pi^2}F_{\mu\nu}F^{\mu\nu},
\end{equation}
which originates from the quantum correction of the electromagnetic fields.

In  hydrodynamics, we usually express the stress tensor in the Landau frame. In Appendix \ref{sec:Landau},
 the symmetric part of the stress tensor  is written in the Landau frame.

\section{Conservation laws}
\label{sec:cons}
With   $j^\mu$ in (\ref{j-0-a}),(\ref{j-1-a}) and (\ref{jv-2}), $j_5^\mu$ in (\ref{j5-0-a}),(\ref{j5-1-a}) and (\ref{j5-2}), and $T^{\mu\nu}$ in (\ref{T0}), (\ref{T1}), (\ref{tvv2-munu}),(\ref{tve2-munu}) and (\ref{tee2-munu-re}),  we can check
 conservation laws for these quantities. In doing so, we must restrict ourselves
to a specific system in constant and homogeneous electromagnetic fields with  conditions (\ref{pd1}),(\ref{pd3}),(\ref{pd2}),
and (\ref{integrability}) or (\ref{integrability-1}). Here we will not present a detailed   derivation, but give
the necessary identities in performing the calculation. These identities hold only under   specific conditions that
are imposed in this paper,
\begin{eqnarray}
\partial_\mu \frac{u^\mu}{T} &=& 0,\ \ \ u\cdot \partial\frac{1}{T} = 0,\ \ \partial_\mu u^\mu = 0,\ \ \
\partial_\mu \frac{1}{T} = -u\cdot \partial\frac{u_\mu}{T} = \frac{\varepsilon_\mu}{T}, \\
\partial_\mu \omega_\nu &=&  \varepsilon\cdot \omega\, g_{\mu\nu} - 2\varepsilon_\mu \omega_\nu,\\
\partial_\mu \varepsilon_\nu &=& \omega_\mu \omega_\nu -\varepsilon_\mu \varepsilon_\nu + \varepsilon^2 u_\mu u_\nu
-\omega^2  \Delta_{\mu\nu}+ \left(u_\mu\epsilon_{\nu\lambda\rho\sigma}+u_\nu \epsilon_{\mu\lambda\rho\sigma}\right)
u^\lambda \varepsilon^\rho \omega^\sigma,\\
\partial_\mu B_\nu &=&-  E_\mu \omega_\nu
 + \varepsilon\cdot B\, u_\mu u_\nu + \omega\cdot E\, \Delta_{\mu\nu}
-\left(u_\mu \epsilon_{\nu\lambda\rho\sigma}+ u_\nu \epsilon_{\mu\lambda\rho\sigma}\right)
u^\lambda \varepsilon^\rho E^\sigma,\\
\partial_\mu E_\nu &=&  B_\mu \omega_\nu
+ \varepsilon\cdot E\, u_\mu u_\nu - \omega\cdot B\, \Delta_{\mu\nu}
+\left(u_\mu \epsilon_{\nu\lambda\rho\sigma}+ u_\nu \epsilon_{\mu\lambda\rho\sigma}\right)
u^\lambda  E^\rho \omega^\sigma.
\end{eqnarray}
With the help of these identities, we can verify  following conservation laws,
\begin{eqnarray}
\partial^\mu j_\mu = 0,\ \
\partial^\mu j_\mu^5 =-\frac{1}{2\pi^2}E\cdot B,\ \
\partial^\mu T_{\mu\nu} = F_{\nu\mu}j^\mu.
\end{eqnarray}
We note  that the second-order correction to the axial current does not contribute to the chiral anomaly as it should be.
We find that the term proportional to $\ln\Lambda/T$ in Eq.\,(\ref{tee2-munu-re})
is essential to conserve the energy-momentum  when the vorticity is present.
We can seperate the energy-momentum tensor into a symmetric and an antisymmetric part,
\begin{eqnarray}
T^{\mu\nu} = T_S^{\mu\nu} +  T_A^{\mu\nu},
\end{eqnarray}
where the symmetric and antisymmetric part   are given by,
\begin{eqnarray}
T^{\mu\nu}_S&=& \rho  u^\mu u^\nu -\frac{1}{3}\rho\Delta^{\mu\nu} + n_5 \left(u^\mu \omega^\nu + u^\nu \omega^\mu \right)
+\frac{\xi}{2}\left(u^\mu B^\nu + u^\nu B^\mu \right)  \nonumber\\
& &-\frac{1}{2}\xi_5 \left[{3 u^\mu u^\nu ( \omega^2+\varepsilon^2)  -\Delta^{\mu\nu} ( \omega^2+\varepsilon^2)}
 -2( u^\mu \epsilon^{\nu\alpha\beta\gamma} + u^\nu \epsilon^{\mu\alpha\beta\gamma} ) u_\alpha \varepsilon_\beta \omega_\gamma \right]\nonumber\\
& &-\frac{1}{2}\xi_{B5} \left[u^\mu u^\nu (\omega\cdot B+\varepsilon\cdot E)
- (\omega^\mu B^\nu + E^\mu \varepsilon^\nu)
-(u^\mu \epsilon^{\nu\alpha\beta\gamma}+u^\nu \epsilon^{\mu\alpha\beta\gamma})u_\alpha E_\beta \omega_\gamma  \right]\nonumber\\
& &+ \kappa_1^E u^\mu u^\nu E^2 +\kappa_1^B u^\mu u^\nu B^2
+\kappa_2^E \Delta^{\mu\nu} E^2 + \kappa_2^B\Delta^{\mu\nu} B^2
+\kappa_3\left( E^\mu E^\nu +B^\mu B^\nu \right)\nonumber\\
& &+\kappa_4\left( u^\mu \epsilon^{\nu\alpha\beta\gamma}
+ u^\nu \epsilon^{\mu\alpha\beta\gamma}   \right)u_\alpha E_\beta B_\gamma ,
\end{eqnarray}
\begin{eqnarray}
T^{\mu\nu}_A &=&-\frac{ n_5}{2}\left(u^\mu \omega^\nu - u^\nu \omega^\mu +\epsilon^{\mu\nu\alpha\beta}u_\alpha \varepsilon_\beta \right)
-\frac{\xi}{2} \epsilon^{\mu\nu\alpha\beta}u_\alpha E_\beta
+\xi_5  (u^\mu \epsilon^{\nu\alpha\beta\gamma} - u^\nu \epsilon^{\mu\alpha\beta\gamma})  u_\alpha \varepsilon_\beta \omega_\gamma \nonumber\\
& &+\xi_{B5}  ( u^\mu \epsilon^{\nu\alpha\beta\gamma} -   u^\nu \epsilon^{\mu\alpha\beta\gamma} )u_\alpha E_\beta \omega_\gamma
+\frac{1}{8\pi^2}\left(u^\mu \epsilon^{\nu\alpha\beta\gamma}
- u^\nu \epsilon^{\mu\alpha\beta\gamma}   \right)u_\alpha E_\beta B_\gamma,
\end{eqnarray}
where we have expressed the energy-momentum tensor  in terms of $E_\mu$ and $B_\mu$
and used the renormalized result in Eq.~(\ref{tee2-munu-re}). The coefficients
are defined as
\begin{eqnarray}
\kappa_1^E&=&-\frac{1}{12\pi^2}\left(\hat\kappa +  \ln \frac{\Lambda}{T}-\frac{1}{2}\right),\ \
\kappa_1^B = -\frac{1}{12\pi^2}\left(\hat\kappa+ \ln \frac{\Lambda}{T}\right),\nonumber\\
\kappa_2^E&=&\frac{1}{12\pi^2}\left(\hat\kappa+ \ln \frac{\Lambda}{T}-\frac{1}{2}\right),\ \ \ \
\kappa_2^B = \frac{1}{12\pi^2}\left(\hat\kappa+ \ln \frac{\Lambda}{T}-1\right),\nonumber\\
\kappa_3&=&-\frac{1}{6\pi^2}\left(\hat\kappa+ \ln \frac{\Lambda}{T}-1\right),\ \ \ \
\kappa_4 =- \frac{1}{6\pi^2}\left(\hat\kappa+ \ln \frac{\Lambda}{T}-\frac{3}{4}\right).
\end{eqnarray}
We can verify  following conservation equations,
\begin{eqnarray}
\partial_\mu T^{\mu\nu}_S = F^{\nu\mu}j_\mu,\ \ \ \ \partial_\mu T^{\mu\nu}_A = 0.
\end{eqnarray}

\section{General solution  }
\label{sec:Extend}
We have  presented a specific solution in the previous sections by setting  $f^{(1)}=f^{(2)}=0$ in Eq.(\ref{calJ}).  The general solution under global equilibrium conditions (\ref{pd1}-\ref{pd2}) should be
 a summation of this specific solution and all  possible contributions associated with  nonvanishing $f^{(1)}$ and $f^{(2)}$.  In this section, we determine these contributions. We note that these terms are proportional to $p^\mu \delta(p^2)$ and automatically
satisfy  Eq.~(\ref{Js-c1-n}). They  can not be further constrained by using Eq.~(\ref{Js-c2-n})  because they do not appear on the left-hand side of the equation
due to cancellation.  Hence the remaining constraint can be only from Eq.~(\ref{Js-eq-n}).  From Lorentz invariance together with
charge and  parity invariance, the general  $f^{(1)}$ and $f^{(2)}$ must take following forms,
\begin{eqnarray}
\label{X-1-f}
f^{(1)}
&=& (\omega\cdot p)\beta^2  \mathcal{X}^{(1)}_\omega + (B\cdot p) \beta^3 \mathcal{X}^{(1)}_B, \\
\label{X-2-f}
f^{(2)}
&=& \mathscr{X}_{\Omega\Omega}
 + \mathscr{X}_{\Omega F} + \mathscr{X}_{FF},
\end{eqnarray}
where
\begin{eqnarray}
\mathscr{X}_{\Omega\Omega}
&=& \omega^2 \beta^2 \mathcal{X}^{(2)}_{\omega\omega1}  + \varepsilon^2 \beta^2 \mathcal{X}^{(2)}_{\varepsilon\varepsilon1}
+ (\omega\cdot p)^2 \beta^4  \mathcal{X}^{(2)}_{\omega\omega2} + (\varepsilon\cdot p)^2 \beta^4 \mathcal{X}^{(2)}_{\varepsilon\varepsilon2}\nonumber\\
& & +\epsilon^{\nu\lambda\rho\sigma}u_\nu p_\lambda\omega_\rho \varepsilon_\sigma   \beta^3 \mathcal{X}^{(2)}_{\omega\varepsilon},\\
\mathscr{X}_{\Omega F}
&=& \omega\cdot B \beta^3 \mathcal{X}^{(2)}_{\omega B 1}  + \varepsilon\cdot E \beta^3 \mathcal{X}^{(2)}_{\varepsilon E 1}
+ (\omega\cdot p)(B\cdot p) \beta^5  \mathcal{X}^{(2)}_{\omega B2}
+ (\varepsilon\cdot p)(E\cdot p) \beta^5 \mathcal{X}^{(2)}_{\varepsilon E2}\nonumber\\
& & +\epsilon^{\nu\lambda\rho\sigma}u_\nu p_\lambda\omega_\rho E_\sigma   \beta^4 \mathcal{X}^{(2)}_{\omega E},\\
\mathscr{X}_{FF}
&=& B^2 \beta^4 \mathcal{X}^{(2)}_{BB1}  + E^2 \beta^4 \mathcal{X}^{(2)}_{EE1}
+ (B\cdot p)^2\beta^6  \mathcal{X}^{(2)}_{BB2} + (E\cdot p)^2 \beta^6 \mathcal{X}^{(2)}_{EE2}\nonumber\\
& & +\epsilon^{\nu\lambda\rho\sigma}u_\nu p_\lambda B_\rho E_\sigma   \beta^5 \mathcal{X}^{(2)}_{BE},
\end{eqnarray}
where all  $\mathcal{X}$ functions depends on variables
\begin{eqnarray}
z= \beta\cdot p -\bar \mu_s,\ \ \ \ \tilde z = \beta\cdot p +\bar \mu_s.
\end{eqnarray}
From Eq.~(\ref{Js-eq-n}) at the first order, we have
\begin{eqnarray}
\label{eq-pf1}
0&=&\nabla^\mu \left[ p_\mu f^{(1)}\delta(p^2)\right]\nonumber\\
&=&\delta(p^2)\left[
 3 (B\cdot p)(\varepsilon\cdot p)
 -( E\cdot p) (\omega\cdot p)
  + (\varepsilon\cdot B) (u\cdot p)^2 \right.\nonumber\\
& &\hspace{1cm}\left.
 + (\omega\cdot E) \bar p^2 -2(u\cdot p) \epsilon_{\nu\lambda\rho\sigma}p^\nu u^\lambda \varepsilon^\rho E^\sigma \right]
 \beta^3 \mathcal{X}^{(1)}_B\nonumber\\
& &+\delta(p^2)\left[ (\omega\cdot E)(u\cdot p)\beta^2 \mathcal{X}^{(1)}_\omega
+ \epsilon_{\mu\nu\rho\sigma}u^\mu p^\nu \omega^\rho B^\sigma \beta^2 \mathcal{X}^{(1)}_\omega
+( E\cdot B) (u\cdot p)\beta^3 \mathcal{X}^{(1)}_B\right]\nonumber\\
& &-\delta(p^2)  \left[ (\omega\cdot p)\beta^2 \partial_{\tilde z} \mathcal{X}^{(1)}_\omega
+ (B\cdot p) \beta^4 \partial_{\tilde z} \mathcal{X}^{(1)}_B  \right]2\beta (E\cdot p).
\end{eqnarray}
Obviously, this equation holds only if
\begin{eqnarray}
\mathcal{X}^{(1)}_B = \mathcal{X}^{(1)}_\omega =0,
\end{eqnarray}
which means that there is no  solution at the first order with nonvanishing $f^{(1)}$. However, it is interesting to note that when we  turn off the electromagnetic field at the beginning, Eq.~(\ref{eq-pf1}) holds automatically, so we have
\begin{eqnarray}
\label{X-1-f}
f^{(1)} &=& (\omega\cdot p)\beta^2  \mathcal{X}^{(1)}_\omega.
\end{eqnarray}
 Hence the first order solution  in Eq.(\ref{Jmu-1-a}) is
a unique solution at global equilibrium. Similarly, Eq.~(\ref{Js-eq-n}) at the second order is,
\begin{eqnarray}
0&=&\nabla^\mu \left[ p_\mu f^{(2)}\delta(p^2)\right]\nonumber\\
&=&\delta(p^2)p_\mu \nabla^\mu \left[ \mathscr{X}_{\Omega\Omega}
 + \mathscr{X}_{\Omega F} + \mathscr{X}_{FF} \right].
\end{eqnarray}
We find that this equation holds only if  following relations are fulfilled,
\begin{eqnarray}
\mathcal{X}^{(2)}_{BB1} &=& \mathcal{X}^{(2)}_{EE1}
= \mathcal{X}^{(2)}_{BB2} = \mathcal{X}^{(2)}_{EE2}
= \mathcal{X}^{(2)}_{BE2} =0,\\
\mathcal{X}^{(2)}_{\omega B1} &=&\mathcal{X}^{(2)}_{\omega B2}
=\mathcal{X}^{(2)}_{\varepsilon E2} =\mathcal{X}^{(2)}_{\omega E} = 0,\\
{X}^{(2)}_{\omega\omega2}&=&{X}^{(2)}_{\varepsilon\varepsilon2}=0,\\
 \mathcal{X}^{(2)}_{\varepsilon E 1} &=& \frac{1}{2}  \mathcal{X}^{(2)}_{\omega\varepsilon},\\
\end{eqnarray}
together with
\begin{eqnarray}
\label{Xoe1}
\mathcal{X}^{(2)}_{\omega\varepsilon} &=&\frac{1}{\beta\cdot p}
\left( \mathcal{X}^{(2)}_{\omega\omega1} + \mathcal{X}^{(2)}_{\varepsilon\varepsilon1}\right),\\
\label{Xoe2}
 \mathcal{X}^{(2)}_{\omega\varepsilon} &=& 2\partial_{\tilde z}\mathcal{X}^{(2)}_{\varepsilon\varepsilon1},\\
 \label{Xoe3}
\partial_{\tilde z}  \mathcal{X}^{(2)}_{\omega\varepsilon } &=& \partial_{\tilde z} \mathcal{X}^{(2)}_{\omega\omega1} =0.
 \end{eqnarray}
It is easy to verify that the general solution must take the form of
\begin{eqnarray}
\mathcal{X}^{(2)}_{\omega\omega1}&=& a(z),\\
\mathcal{X}^{(2)}_{\omega\varepsilon } &=& b(z), \\
\mathcal{X}^{(2)}_{\varepsilon\varepsilon1}&=&  (\beta\cdot p) b(z) - a(z).
 \end{eqnarray}
where $a(z)$ and $b(z)$ are  arbitrary real functions of $z$.
Hence the general form of $f^{(2)}$ is
\begin{eqnarray}
f^{(2)}
&=&\omega^2\beta^2 a(z) + \varepsilon^2\beta^2 \left[(\beta\cdot p) b(z) -a(z)\right]\nonumber\\
& &+ \epsilon^{\nu\lambda\rho\sigma}u_\nu p_\lambda\omega_\rho \varepsilon_\sigma   \beta^3 b(z)
+\frac{1}{2}(\varepsilon\cdot E) \beta^3 b(z).
\end{eqnarray}
We see that there is no  contribution from electromagnetic fields except a $\varepsilon\cdot E$ term.
Integrating over  momenta, we obtain contributions to the current and stress tensor
\begin{eqnarray}
\Delta j^{(2)}_\mu &=& \int d^4 p f^{(2)} p_\mu \delta(p^2)\nonumber\\
&=& \left[A_j \omega^2  + B_j \varepsilon^2  + C_j (\varepsilon\cdot E)\right] u_\mu
+ D_j \epsilon_{\mu \nu \rho\sigma} u^\nu \omega^\rho \varepsilon^\sigma , \\
\Delta T^{(2)}_{\mu\nu} &=& \int d^4 p f^{(2)} p_\mu p_\nu \delta(p^2)\nonumber\\
&=& \left[ A_T \omega^2   + B_T \varepsilon^2  + C_T (\varepsilon\cdot E) \right] \left(u_\mu u_\nu -\frac{1}{3}\Delta_{\mu\nu}\right)\nonumber\\
& & \hspace{0.2cm}+ D_T \left( u_\mu \epsilon_{\nu \lambda \rho\sigma}  +u_\nu \epsilon_{\mu \lambda \rho\sigma} \right) u^\lambda \omega^\rho \varepsilon^\sigma,
\end{eqnarray}
where we have used $\Delta$ to denote these are additional contribution from nonvanishing $f^{(2)}$. After integration over $p_0$,
these coefficients are
\begin{eqnarray}
A_j &=& \frac{1}{2}\beta^2\int d^3 p \left[ a(z_+) - a(z_-)\right], \\
B_j &=& \frac{1}{2}\beta^2\int d^3 p \left\{\beta p \left[ b(z_+) +  b(z_- )\right] -\left[ a(z_+) - a(z_-)\right]\frac{}{} \right\} ,\\
C_j &=& \frac{1}{4}\beta^3 \int d^3 p  \left[ b(z_+) - b(z_-)\right], \\
D_j &=& \frac{1}{6} \beta^3 \int d^3 p \, p \left[ b(z_+) + b(z_-)\right],\\
A_T &=& \frac{1}{2}\beta^2\int d^3 p \, p \left[ a(z_+) + a(z_-)\right], \\
B_T &=& \frac{1}{2}\beta^2\int d^3 p \, p \left\{\beta p \left[ b(z_+) -  b(z_- )\right] -\left[ a(z_+) + a(z_-)\right]\frac{}{} \right\}, \\
C_T &=& \frac{1}{4}\beta^3 \int d^3 p \, p  \left[ b(z_+) + b(z_-)\right], \\
D_T &=& \frac{1}{6} \beta^3 \int d^3 p \, p^2 \left[ b(z_+) - b(z_-)\right],
\end{eqnarray}
where $z_\pm = \beta \cdot p \mp \bar\mu_s$.
Although  unknown funcitons $a(z)$ and $b(z)$ are  arbitrary, only two of these coefficients are  independent. It is convenient to choose $A_T$ and $D_T$ as  independent variables from which other coefficients can be derived through  following relations
\begin{eqnarray}
\label{Aj}
A_j &=&\frac{1}{3\beta}\frac{\partial (\beta^2 A_T)}{\partial\bar\mu_s}, \\
D_j &=&\frac{1}{4\beta}\frac{\partial (\beta^2 D_T)}{\partial\bar\mu_s}, \\
B_j &=& 3 D_j - A_j, \\
C_j &=&\frac{1}{2} \frac{\partial (\beta D_j)}{\partial\bar\mu_s},  \\
C_T &=&\frac{3}{2}D_j, \\
\label{BT}
B_T &=&3 D_T - A_T.
\end{eqnarray}
In Refs.(\cite{Buzzegoli:2017cqy},\cite{Buzzegoli:2018wpy}), the authors calculate the second-order transport coefficients in global equilibrium without
electromagnetic fields. We find that their results  are different from our results in Eqs.(\ref{jv-2},\ref{j5-2},\ref{tvv2-munu}) from our special  solution
(\ref{Jmu-2-a}) with vanishing $f^{(1)}$ and $f^{(2)}$.
The difference between our result and the result in  Refs.\cite{Buzzegoli:2017cqy,Buzzegoli:2018wpy}  are given by
 \begin{eqnarray}
\label{dj-B}
\Delta j^{\mu}_s
&=& \frac{\bar\mu_s}{8\pi^2\beta} ( \varepsilon^2+\omega^2 )u^\mu
+\frac{\bar\mu_s}{12\pi^2\beta}\epsilon^{\mu\alpha\beta\gamma} u_\alpha \varepsilon_\beta \omega_\gamma,\\
\label{T-B}
\Delta T^{\mu\nu}_s &=& \frac{\beta}{2}s\xi_s
\left[\left(3 u^\mu u^\nu - \Delta^{\mu\nu}\right)\left(\frac{1}{2} \omega^2+ \frac{5}{6}\varepsilon^2\right)
+\frac{4}{3}( u^\mu \epsilon^{\nu\alpha\beta\gamma} + u^\nu \epsilon^{\mu\alpha\beta\gamma} )
u_\alpha  \omega_\beta \varepsilon_\gamma\right],
\end{eqnarray}
with $\xi_s$ being given in Eq.~(\ref{xis}).   From these expressions, we can read off
\begin{eqnarray}
A_T&=&\frac{1 }{16\pi^2\beta^2}\left[\pi^2+3\bar\mu_s^2\frac{}{}\right],\\
D_T&=& \frac{1 }{18\pi^2\beta^2}\left[\pi^2+3\bar\mu_s^2\frac{}{}\right].
\end{eqnarray}
Substituting it into  the relations (\ref{Aj}-\ref{BT}), {we can obtain all other coefficients } which are exactly consistent with
the result of Refs.\cite{Buzzegoli:2017cqy,Buzzegoli:2018wpy} as well as   $C_j = {1 }/{24\pi^2},C_T = {\mu_s }/{8\pi^2}$ for
$\varepsilon\cdot E$  terms when  electromagnetic fields are involved.  
This  implies  that both results, ours and that of Refs.\cite{Buzzegoli:2017cqy,Buzzegoli:2018wpy},  are  possible solutions which should correspond to two different
density matrices.

\section{Summary and Discussion}
\label{sec:summary}
We have derived in the covariant Wigner function formalism  the charge and
chiral  currents as well as the stress tensor for chiral fermions in  uniform
 vorticity and electromagnetic fields up to the second order of spatial derivatives.
We present all possible second order contributions in quadratic forms of the vorticity and electromagnetic field.
These contributions include coupling terms of electromagnetic-field-electromagnetic-field (ee),
vorticity-vorticity (vv) and vorticity-electromagnetic-field (ve).
All the  terms can modify the charge density, while only the `ee' and `vv' terms
can modify the chiral charge density and `ve' term can not contribute to it. We find that the electromagnetic field
can induce Hall currents in the form $\epsilon^{\mu\nu\rho\sigma} u_\nu  E_\rho B_\sigma$ in the charge and axial charge current .
There is also a Hall term $\epsilon^{\mu\nu\rho\sigma} u_\nu  E_\rho \omega_\sigma$ in the charge current.
For the energy-momentum tensor at the second order, we find that the vorticity and electromagnetic  field
contribute to the energy density and pressure. The conservation laws as well as chiral and trace anomaly can be verified
with our second-order solution. All coefficients we obtain in this work can be directly
applied to the anomalous hydrodynamics as inputs. We also demonstrate that the solution in global equilibrium is fully constrained  at the first order
while solutions at the second order can only be constrained up to some unknown functions. These
unknown functions appear in  the `vv' part and $\varepsilon\cdot E$ term in the  `ve' part.
Other contributions can be fully constrained.

{In this work, we restrict ourselves to a non-interacting chiral system without collision terms under the constant electromagnetic and vorticity field. If we go beyond to include collision terms but still in global equilibrium, collision terms won't change present results since collision terms always vanish in global equilibrium. Instead, if we consider non-equilibrium and varying fields, our present results still provide a baseline for studying these effects. We can expand the solution around the results in global equilibrium given here and investigate contributions from collisions or variation of fields. We reserve the topics along this line in a future study.}

\acknowledgments

JHG would like to thank Shu Lin for helpful discussion. This work was supported in part by the National Natural Science Foundation of China under grant
No. 11890713, 11675092, 11535012 and 11947301, and the Natural Science Foundation of Shandong Province under No. JQ201601.

\appendix

\section{Integration in $T_{s,{\textrm{ee}}}^{(2)\mu\nu}$ }
\label{sec:Integation}
In this Appendix, we  give detailed derivations of some integrals in  $T_{s,{\textrm{ee}}}^{(2)\mu\nu}$.
All other coefficients in $j_s^{(2)}$, $T_{s,{\textrm{vv}}}^{(2)\mu\nu}$ and $T_{s,{\textrm{ve}}}^{(2)\mu\nu}$
 can be derived in a similar way. We   take the integral in the first term of Eq.~(\ref{T-s-ee-1}) as an example,
\begin{eqnarray}
I&\equiv&\int d^{4-\epsilon} p\ (u\cdot p)^4 f \delta'''(p^2),
\end{eqnarray}
 First, we need to integrate out the $\delta$ function by using the identity
\begin{eqnarray}
\delta'''\left(p^2\right)&=&\frac{1}{16 p_0|\vec p|^3}\left[\frac{}{}\delta'''\left(p_0-|\vec p|\right)+ \delta'''\left(p_0+|\vec p|\right)\right]\nonumber\\
& &+\frac{3}{16 p_0|\vec p|^4}\left[\frac{}{}\delta''\left(p_0-|\vec p|\right)-\delta''\left(p_0+|\vec p|\right)\right]\nonumber\\
& &+\frac{3}{16p_0|\vec p|^5}\left[\frac{}{}\delta'\left(p_0-|\vec p|\right)+\delta'\left(p_0+|\vec p|\right)\right],
\end{eqnarray}
and obtain
\begin{eqnarray}
I&=&-\frac{\pi^{(3-\epsilon)/{2}}}{8\Gamma\left((3-\epsilon)/{2}\right)}
\int \frac{d |\vec p|}{|\vec p|^\epsilon} \left[\frac{1}{|\vec p|}\frac{d^3}{d|\vec p|^3}
- \frac{3}{|\vec p|^2}\frac{d^2}{d |\vec p|^2}
+\frac{3}{|\vec p|^3}\frac{d}{d |\vec p| }  \right] \left[|\vec p|^3(f^++f^-)\right] \nonumber\\
&=&-\frac{\pi^{(3-\epsilon)/{2}}}{8\Gamma\left((3-\epsilon)/{2}\right)}
\int \frac{d |\vec p|}{|\vec p|^\epsilon} \left[|\vec p|^2 \frac{d^3}{d|\vec p|^3}
+6|\vec p|\frac{d^2}{d|\vec p|^2} + 3 \frac{d}{d|\vec p|}- 3\frac{1}{|\vec p|}  \right] (f^++f^-),
\end{eqnarray}
where the prefactor is from the integration over the solid angle in $4-\epsilon$ dimension and $f^+/f^-$ denotes
the contribution from $p_0>0/p_0<0$ in Eq.~(\ref{f-p})/(\ref{f-m}).  After integration by parts, we have
\begin{eqnarray}
I&=& \frac{3\pi^{(3-\epsilon)/{2}}}{8\Gamma\left((3-\epsilon)/{2}\right)}
\int d |\vec p| \left[{\frac{1-3\epsilon}{3 |\vec p|^\epsilon }}\frac{d}{d |\vec p|}+\frac{1}{|\vec p|^{1+\epsilon}}\right] (f^++f^-) \nonumber\\
&=& \frac{3\pi^{(3-\epsilon)/{2}}}{8\Gamma\left((3-\epsilon)/{2}\right)}\left(\frac{\epsilon}{3} + 1\right)
\int d |\vec p| \frac{1}{|\vec p|^{1+\epsilon}} (f^++f^-)\nonumber\\
&\equiv&\frac{\left( 3 + \epsilon \right)}{16}\kappa^\epsilon_s,
\end{eqnarray}
where
\begin{eqnarray}
\kappa_s^\epsilon&=&
\frac{2\pi^{(3-\epsilon)/{2}}}{\Gamma\left((3-\epsilon)/{2}\right)}
\int  \frac{d |\vec p|}{|\vec p|^{1+\epsilon}}(f^+ + f^-)\nonumber\\
&=&\frac{4\pi^{(3-\epsilon)/{2}} T^{{-\epsilon}}}{\Gamma\left((3-\epsilon)/{2}\right)(2\pi)^{3-\epsilon}}
\int_0^\infty  \frac{dy}{y^{1+\epsilon}}\left[\frac{1}{e^{y-\bar \mu_s}+1}+\frac{1}{e^{y+\bar\mu_s}+1} {-1}\right].
\end{eqnarray}
It should be noted that if  we did not include the vacuum contribution $-1$ in the square brackets of the last line, the integral would have
infrared divergence. Once we include it as  above, the infrared divergence is cancelled with only the ultraviolet divergence left.
We can single out the divergence by  another  integration by parts
\begin{eqnarray}
\kappa_s^\epsilon&=&
-\frac{4\pi^{\frac{3-\epsilon}{2}} T^{{-\epsilon}} }{\Gamma\left(\frac{3-\epsilon}{2}\right)(2\pi)^{3-\epsilon}}
\left\{\frac{1}{\epsilon}
- \int_0^\infty  dy \ln y\ \left[\frac{e^{y-\bar\mu_s}}{\left(e^{y-\bar\mu_s}+1\right)^2}+\frac{e^{y+\bar\mu_s}}{\left( e^{y+\bar\mu_s}+1\right)^2} \right] \right\}.
\end{eqnarray}
The $1/\epsilon$ pole term corresponds to a logarithmic divergence in the momentum integral.

\section{Results in the Landau frame}

\label{sec:Landau}
In relativistic hydrodynamics, one has a freedom to choose any frame characterized by a different fluid velocity.
The Landau frame is the one in which the fluid velocity satisfies $u_\mu T^{\mu\nu}=\rho u^\nu$.
Since the energy momentum tensor in Sec.~\ref{sec:tensor} has an anti-symmetric part in the first and second order,
in this section, we will rewrite the symmetric part of the stress tensor up
to the second order in the Landau frame. Let us introduce the fluid velocity $U^\mu$ in the Landau frame given by
\begin{eqnarray}
U^\mu &=& u^\mu +\frac{n_5  }{\rho +P}\omega^\mu +\frac{\xi}{2(\rho+P)} B^\mu \nonumber\\
& &-\left[\frac{n_5^2 }{2(\rho+P)^2}\omega^2
+ \frac{\xi^2}{8(\rho+P)^2}B^2
+\frac{n_5 \xi}{2(\rho+P)^2}\omega\cdot B \right] u^\mu\nonumber\\
& &+\frac{\xi_5 }{\rho +P}\epsilon^{\mu\alpha\beta\gamma}u_\alpha\varepsilon_\beta\omega_\gamma
+\frac{\xi_{B5}}{2(\rho+P)}\epsilon^{\mu\alpha\beta\gamma}u_\alpha E_\beta\omega_\gamma
+\frac{\kappa_4}{\rho+P }\epsilon^{\mu\alpha\beta\gamma}u_\alpha E_\beta B_\gamma.
\end{eqnarray}
It is easy to verify that $U^2=1$ up to the second order. From this relation, we can also express $u^\mu$ in terms of $U^\mu$,
\begin{eqnarray}
u^\mu
&=& U^\mu -\frac{n_5  }{\rho +P}\omega^\mu_U - \frac{\xi}{2(\rho+P)} B^\mu_U \nonumber\\
& &-\left[\frac{n_5^2}{2(\rho+P)^2}\omega^2_U + \frac{\xi^2}{8(\rho+P)^2}B^2_U
+\frac{n_5 \xi}{2(\rho+P)^2}\omega_U\cdot B_U\right] U^\mu \nonumber\\
& &-\left[\frac{\xi_5 }{\rho +P}+\frac{n_5^2}{(\rho+P)^2}\right]
\epsilon^{\mu\alpha\beta\gamma}U_\alpha \varepsilon_{U\beta}\omega_{U\gamma}
-\left[\frac{\xi_{B5}}{2(\rho+P)}+\frac{n_5\xi}{(\rho+P)^2}\right]
\epsilon^{\mu\alpha\beta\gamma}U_\alpha E_{U\beta}\omega_{U\gamma}\nonumber\\
& &-\left[\frac{\kappa_4}{\rho+P }+\frac{\xi^2}{4(\rho+P)^2}\right]
\epsilon^{\mu\alpha\beta\gamma}U_\alpha E_{U\beta} B_{U\gamma},
\end{eqnarray}
where $\omega^\mu_U= T\tilde\Omega^{\mu\nu}U_\nu$, $\varepsilon^\mu_U = T \Omega^{\mu\nu}U_\nu$,
$B^\mu_U = \tilde F^{\mu\nu} U_\nu$, and $ E^\mu_U=F^{\mu\nu}U_\nu$ are counterparts of
$\omega^\mu$, $\varepsilon^\mu$, $B^\mu$, and $ E^\mu $  in the Landau frame, respectively.
Up to the second order, the corresponding relations between two groups of quantities are given by
\begin{eqnarray}
\omega^\mu
&=&\omega^\mu_U +\left(\frac{n_5  }{\rho +P}\omega_U^2 + \frac{\xi}{2(\rho+P)} \omega_U\cdot B_U \right) U^\mu\nonumber\\
& &+\frac{n_5  }{\rho +P}\epsilon^{\mu\nu\alpha\beta}\omega_{U\nu} U_\alpha \varepsilon_{U\beta}
+\frac{\xi}{2(\rho+P)}\epsilon^{\mu\nu\alpha\beta}B_{U\nu} U_\alpha \varepsilon_{U\beta},\\
\varepsilon^\mu
&=&\varepsilon^\mu_U +\left(\frac{n_5  }{\rho +P}\varepsilon_U\cdot \omega_U + \frac{\xi}{2(\rho+P)} \varepsilon_U\cdot B_U \right) U^\mu\nonumber\\
& &-\frac{\xi}{2(\rho+P)}\epsilon^{\mu\nu\alpha\beta}B_{U\nu} U_\alpha \omega_{U\beta},\\
B^\mu
&=&B^\mu_U + \left(\frac{n_5  }{\rho +P}\omega_U\cdot B_U + \frac{\xi}{2(\rho+P)}  B_U^2 \right) U^\mu\nonumber\\
& &+\frac{n_5  }{\rho +P}\epsilon^{\mu\nu\alpha\beta}\omega_{U\nu} U_\alpha E_{U\beta}
+\frac{\xi}{2(\rho+P)} \epsilon^{\mu\nu\alpha\beta}B_{U\nu} U_\alpha E_{U\beta},\\
E^\mu
&=&E^\mu_U +\left(\frac{n_5  }{\rho +P}E_U\cdot \omega_U + \frac{\xi}{2(\rho+P)}E_U\cdot  B_U \right) U^\mu\nonumber\\
& &-\frac{n_5 }{\rho +P}\epsilon^{\mu\nu\alpha\beta}\omega_{U\nu} U_\alpha B_{U\beta}.
\end{eqnarray}
Hence the difference between frames  arises only at the second order.
With these equations, we can obtain the symmetric stress tensor in the Landau frame
\begin{eqnarray}
T^{\mu\nu}&=&\left\{ \rho   +\left(\frac{n_5^2 }{\rho+P} - \frac{ 3}{2}\xi_5\right)\omega^2_U
+\left[ \frac{\xi^2}{4(\rho+P)} +\kappa_1^B\right] B^2_U  \right.\nonumber\\
& &\left. + \left(\frac{n_5 \xi}{\rho+P} -\frac{\xi_{B5}}{2}\right)\omega_U\cdot B_U
- \frac{ 3}{2}\xi_5 \varepsilon^2_U
-\frac{1}{2}\xi_{B5} \varepsilon_U\cdot E_U
+\kappa_1^E E^2_U\right\} U^\mu U^\nu \nonumber\\
& &  -\left[ P -\frac{ 1}{2}\xi_5  \left( \omega^2_U+\varepsilon^2_U\right)
-\kappa_2^E E^2_U -  \kappa_2^B  B^2_U\right] \Delta^{\mu\nu}_U\nonumber\\
& &+\left(E^\mu_U E^\nu_U+B^\mu_U B^\nu_U\right) \kappa_3
+ \frac{1}{2}\xi_{B5} \left(\omega^\mu_U B^\nu_U + E^\mu_U \varepsilon^\nu_U\right)\nonumber\\
& &-\frac{n_5^2 }{\rho+P}\omega^\mu_U \omega^\nu_U
-\frac{\xi^2}{4(\rho+P)}B_U^\mu B_U^\nu
-\frac{n_5\xi}{2(\rho+P)}(\omega^\mu_U B_U^\nu +\omega^\nu_U B_U^\mu ),
\end{eqnarray}
where $\Delta^{\mu\nu}_U=g^{\mu\nu}-U^\mu U^\nu$.
The vector current in the Landau frame is given by
\begin{eqnarray}
\label{j-Landau}
j^\mu
 &=& n U^\mu +\left(\xi - \frac{ n n_5  }{\rho +P}\right) \omega^\mu_U +\left(\xi_B- \frac{n\xi}{2(\rho+P)}\right) B^\mu_U \nonumber\\
& &-\left[\frac{n n_5^2 }{2(\rho+P)^2} - \frac{\xi n_5  }{\rho +P}\right]\omega^2_U U^\mu
+\left[ \frac{\xi_B\xi}{2(\rho+P)} - \frac{n \xi^2}{8(\rho+P)^2}\right] B^2_U U^\mu \nonumber\\
& &-\left[\frac{n n_5 \xi}{2(\rho+P)^2}  -\frac{\xi^2}{2(\rho+P)} - \frac{\xi_B n_5  }{\rho +P}\right]\omega_U\cdot B_U U^\mu\nonumber\\
& &- \left(\frac{n\xi_5 }{\rho +P}+\frac{n n_5^2}{(\rho+P)^2} -\frac{\xi n_5  }{\rho +P} \right)
\epsilon^{\mu\alpha\beta\gamma}U_\alpha \varepsilon_{U\beta}\omega_{U\gamma}\nonumber\\
& &+\left[\frac{\xi_B\xi}{2(\rho+P)}-\frac{n\xi^2}{4(\rho+P)^2}-\frac{n\kappa_4}{\rho+P }\right]
\epsilon^{\mu\alpha\beta\gamma}U_\alpha E_{U\beta} B_{U\gamma} \nonumber\\
& &+\left[\frac{\xi^2}{2(\rho+P)}-\frac{n n_5 \xi}{2(\rho+P)^2}\right]
\epsilon^{\mu\alpha\beta\gamma} U_\alpha \varepsilon_{U\beta} B_{U\gamma}\nonumber\\
& &-\left[\frac{n\xi_{B5}}{2(\rho+P)}+\frac{n n_5 \xi}{2(\rho+P)^2}-\frac{\xi_B n_5  }{\rho +P}\right]\epsilon^{\mu\alpha\beta\gamma}U_\alpha E_{U\beta}\omega_{U\gamma}\nonumber\\
& & -   \xi_{B5} ( \varepsilon^2_U+\omega^2_U )U_\mu
-\frac{1}{4\pi^2}\left[  (\omega_U\cdot B_U + \varepsilon_U\cdot E_U)U_\mu
+\epsilon_{\mu \nu \rho\sigma} U^\nu E^\rho_U \omega^\sigma_U\right]\nonumber\\
& &-\frac{C}{12\pi^2} \left[ U_\mu( E^2_U+B^2_U ) + 2\epsilon_{\mu\nu\rho\sigma} U^\nu  E^\rho_U B^\sigma_U \right].
\end{eqnarray}
The axial current in the Landau frame is given by
\begin{eqnarray}
\label{j5-Landau}
j_{5}^\mu
 &=& n_5 U^\mu +\left(\xi_5 -\frac{  n_5^2  }{\rho +P}\right) \omega^\mu_U +\left(\xi_{B5}- \frac{n_5\xi}{2(\rho+P)}\right) B^\mu_U \nonumber\\
& &-\left[\frac{ n_5^3 }{2(\rho+P)^2} - \frac{\xi_5 n_5  }{\rho +P}\right]\omega^2_U U^\mu
+\left[ \frac{\xi_{B5}\xi}{2(\rho+P)} - \frac{n_5 \xi^2}{8(\rho+P)^2}\right] B^2_U U^\mu \nonumber\\
& &-\left[\frac{n_5^2 \xi}{2(\rho+P)^2}  -\frac{\xi_5\xi}{2(\rho+P)} - \frac{\xi_{B5} n_5 }{\rho +P}\right]\omega_U\cdot B_U U^\mu\nonumber\\
& &-\frac{n_5^3}{(\rho+P)^2}\epsilon^{\mu\alpha\beta\gamma}U_\alpha \varepsilon_{U\beta}\omega_{U\gamma}
+\left[\frac{\xi_{B5}\xi}{2(\rho+P)}-\frac{n_5\kappa_4}{\rho+P } -\frac{n_5 \xi^2}{4(\rho+P)^2}\right] \epsilon^{\mu\alpha\beta\gamma}U_\alpha E_{U\beta} B_{U\gamma} \nonumber\\
& &+\left[\frac{\xi_5\xi }{2(\rho+P)}-\frac{n_5^2\xi}{2(\rho+P)^2}\right]
\epsilon^{\mu\alpha\beta\gamma} U_\alpha \varepsilon_{U\beta} B_{U\gamma}
+\left[\frac{n_5\xi_{B5}}{2(\rho+P)}-\frac{n_5^2\xi}{2(\rho+P)^2}\right]
\epsilon^{\mu\alpha\beta\gamma}U_\alpha E_{U\beta}\omega_{U\gamma}\nonumber\\
& &-  \xi_B( \varepsilon^2_U+\omega^2_U ) U^\mu
-\frac{ C_5}{12\pi^2} \left[( E^2_U+B^2_U ) U^\mu + 2\epsilon^{\mu\nu\rho\sigma} U_\nu  E_{U\rho U} B_{U\sigma} \right].
\end{eqnarray}

\end{document}